\documentclass[onecolumn,11pt]{IEEEtran}
\usepackage{soul}
\usepackage[mathscr]{eucal}
\usepackage[cmex10]{amsmath}
\usepackage{epsfig,epsf,psfrag}
\usepackage{amssymb,amsmath,amsthm,amsfonts,latexsym}
\usepackage{amsmath,graphicx,bm,xcolor,url,overpic}
\usepackage[caption=false]{subfig}
\usepackage{fixltx2e}
\usepackage{array}
\usepackage{verbatim}
\usepackage{bm}
\usepackage{algpseudocode}
\usepackage{algorithm}
\usepackage{verbatim}
\usepackage{textcomp}
\usepackage{mathrsfs}
\usepackage{epstopdf}

\newcommand{\openone}{\leavevmode\hbox{\small1\normalsize\kern-.33em1}}

\catcode`~=11 \def\UrlSpecials{\do\~{\kern -.15em\lower .7ex\hbox{~}\kern .04em}} \catcode`~=13

\allowdisplaybreaks[4]

\newcommand{\nn}{\nonumber}

\newcommand{\calA}{\mathcal{A}}
\newcommand{\calB}{\mathcal{B}}
\newcommand{\calC}{\mathcal{C}}
\newcommand{\calD}{\mathcal{D}}
\newcommand{\calE}{\mathcal{E}}
\newcommand{\calF}{\mathcal{F}}
\newcommand{\calG}{\mathcal{G}}

\newcommand{\calM}{\mathcal{M}}

\newcommand{\calP}{\mathcal{P}}

\newcommand{\calS}{\mathcal{S}}
\newcommand{\calT}{\mathcal{T}}

\newcommand{\calW}{\mathcal{W}}
\newcommand{\calX}{\mathcal{X}}

\newcommand{\bQ}{\mathbf{Q}}

\newcommand{\bv}{\mathbf{v}}

\newcommand{\bx}{\mathbf{x}}
\newcommand{\bX}{\mathbf{X}}

\newcommand{\rmA}{\mathrm{A}}

\newcommand{\rmc}{\mathrm{c}}

\newcommand{\rmG}{\mathrm{G}}

\newcommand{\rmH}{\mathrm{H}}

\newcommand{\rmN}{\mathrm{N}}

\newcommand{\rmP}{\mathrm{P}}

\newcommand{\rmr}{\mathrm{r}}

\newcommand{\rmS}{\mathrm{S}}

\newcommand{\rmu}{\mathrm{u}}

\newcommand{\bbN}{\mathbb{N}}

\newcommand{\bbP}{\mathbb{P}}

\newcommand{\bbR}{\mathbb{R}}

\DeclareMathAlphabet{\mathbsf}{OT1}{cmss}{bx}{n}
\DeclareMathAlphabet{\mathssf}{OT1}{cmss}{m}{sl}

\DeclareSymbolFont{bsfletters}{OT1}{cmss}{bx}{n}
\DeclareSymbolFont{ssfletters}{OT1}{cmss}{m}{n}
\DeclareMathSymbol{\bsfGamma}{0}{bsfletters}{'000}
\DeclareMathSymbol{\ssfGamma}{0}{ssfletters}{'000}
\DeclareMathSymbol{\bsfDelta}{0}{bsfletters}{'001}
\DeclareMathSymbol{\ssfDelta}{0}{ssfletters}{'001}
\DeclareMathSymbol{\bsfTheta}{0}{bsfletters}{'002}
\DeclareMathSymbol{\ssfTheta}{0}{ssfletters}{'002}
\DeclareMathSymbol{\bsfLambda}{0}{bsfletters}{'003}
\DeclareMathSymbol{\ssfLambda}{0}{ssfletters}{'003}
\DeclareMathSymbol{\bsfXi}{0}{bsfletters}{'004}
\DeclareMathSymbol{\ssfXi}{0}{ssfletters}{'004}
\DeclareMathSymbol{\bsfPi}{0}{bsfletters}{'005}
\DeclareMathSymbol{\ssfPi}{0}{ssfletters}{'005}
\DeclareMathSymbol{\bsfSigma}{0}{bsfletters}{'006}
\DeclareMathSymbol{\ssfSigma}{0}{ssfletters}{'006}
\DeclareMathSymbol{\bsfUpsilon}{0}{bsfletters}{'007}
\DeclareMathSymbol{\ssfUpsilon}{0}{ssfletters}{'007}
\DeclareMathSymbol{\bsfPhi}{0}{bsfletters}{'010}
\DeclareMathSymbol{\ssfPhi}{0}{ssfletters}{'010}
\DeclareMathSymbol{\bsfPsi}{0}{bsfletters}{'011}
\DeclareMathSymbol{\ssfPsi}{0}{ssfletters}{'011}
\DeclareMathSymbol{\bsfOmega}{0}{bsfletters}{'012}
\DeclareMathSymbol{\ssfOmega}{0}{ssfletters}{'012}

\newcommand{\tilP}{\tilde{P}}

\newcommand{\hatT}{\hat{T}}

\DeclareMathOperator*{\argmax}{arg\,max}
\DeclareMathOperator*{\argmin}{arg\,min}

\newtheorem{theorem}{Theorem}

\newtheorem{definition}{Definition}

\usepackage{graphicx,cite}
\usepackage{epstopdf}
\usepackage{enumerate}
\usepackage{color}
\usepackage{bbm}
\usepackage{booktabs}
\usepackage{makecell}
\usepackage[ colorlinks = true,
linkcolor = blue,
urlcolor  = blue,
citecolor = red,
anchorcolor = green,]{hyperref}

\makeatletter

\newcommand{\Rmnum}[1]{\expandafter\@slowromancap\romannumeral #1@}
\makeatother
\usepackage{setspace}
\def\BibTeX{{\rm B\kern-.05em{\sc i\kern-.025em b}\kern-.08em
T\kern-.1667em\lower.7ex\hbox{E}\kern-.125emX}}

\begin{document}
\title{Exponentially Consistent Low Complexity Tests for Outlier Hypothesis Testing}

\author{Jun Diao, Jingjing Wang and Lin Zhou
\thanks{J. Diao and J. Wang are with the School of Cyber Science and Technology, Beihang University, Beijing, China, 100191, (Emails: \{jundiao,~drwangjj\}@buaa.edu.cn).}
\thanks{L. Zhou is with the School of Automation and Intelligent Manufacturing, Southern University of Science and Technology, Shenzhen, China, 518055 (Email: zhoul9@sustech.edu.cn).}

}
\maketitle

\begin{abstract}
We revisit outlier hypothesis testing, propose exponentially consistent low complexity fixed-length and sequential tests and show that our tests achieve better tradeoff between detection performance and computational complexity than existing tests that use exhaustive search. Specifically, in outlier hypothesis testing, one is given a list of observed sequences, most of which are generated i.i.d. from a nominal distribution while the rest sequences named outliers are generated i.i.d. from another anomalous distribution. The task is to identify all outliers when both the nominal and anomalous distributions are unknown. There are two basic settings: fixed-length and sequential. In the fixed-length setting, the sample size of each observed sequence is fixed a priori while in the sequential setting, the sample size is a random number that can be determined by the test designer to ensure reliable decisions. For the fixed-length setting, we strengthen the results of Bu \emph{et. al} (TSP 2019) by i) allowing for scoring functions beyond KL divergence and further simplifying the test design when the number of outliers is known and ii) proposing a new test, explicitly bounding the detection performance of the test and characterizing the tradeoff among exponential decay rates of three error probabilities when the number of outliers is unknown. For the sequential setting, our tests for both cases are novel and enable us to reveal the benefit of sequentiality. Finally, for both fixed-length and sequential settings, we demonstrate the penalty of not knowing the number of outliers on the detection performance.
\end{abstract}


\section{Introduction}
Outlier hypothesis testing (OHT) is a typical problem in statistical inference, aiming to detect outliers that behave differently from the majority among a given list of sequences. OHT has wide applications across diverse domains including anomaly detection~\cite{zhang2017statistical,gurevich2019sequential,hemo2020searching}, signal detection~\cite{patel2009optimal,wang2009new,brown2000nonparametric}, financial fraud detection~\cite{kong2024asia} and network intrusion detection~\cite{tartakovsky2006novel}. In OHT, one is given a list of observed sequences: the majority named nominal samples are generated i.i.d. from a nominal distribution, while the rest few sequences named outliers are generated i.i.d. from an anomalous distribution. One has no prior knowledge concerning the nominal and anomalous distributions except that the nominal and anomalous distributions are different. The goal of OHT is to design a non-parametric test to identify all the outliers for both cases where the number of outliers is known and unknown.

There are two basic settings: fixed-length and sequential. In the fixed-length setting, the sample size of each observed sequence is fixed a priori while in the sequential setting, the sample size is a random number that can be determined by the test designer to ensure reliable decisions. For the fixed-length setting, Li, Nitinawarat and Veeravalli~\cite[Theorem 8]{li2014} and Zhou, Wei and Hero~\cite[Theorem 5]{zhou2022second} proposed asymptotically optimal tests and characterized the exponential decay rates (error exponents) of various error probabilities. For the sequential setting,  Li, Nitinawarat and Veeravalli~\cite{li2017universal} proposed a non-parametric test that has bounded error probabilities under any pairs of nominal and anomalous distributions and upper bounded the expected sample size. Diao and Zhou~\cite{diao2025sequential} proposed another non-parametric sequential test that has bounded expected stopping time under any pair of nominal and anomalous distributions and characterized the exponential decay rates of error probabilities.

However, all above tests use exhaustive search, which incurs very high computational complexity and renders these tests infeasible for practical applications. For example, when there are $20$ outliers among $100$ observed sequences, if the number of outliers is known, there are $5.36\times 10^{20}$ possibilities concerning the true set of outliers. This number is prohibitively large to run any exhaustive search test.

To address the above problem, for the fixed-length setting, Bu, Zou and Veeravalli~\cite[Algorithm 2 and 3]{bu2019linear} proposed low-complexity tests, proved that their tests are exponentially consistent either when the number of outliers is known or when the number of outliers is unknown but positive. However, there are two limitations of \cite{bu2019linear}. Firstly, when the number of outliers is known, the scoring function is restricted to KL divergence instead of the generalized Jensen-Shannon (GJS) divergence adopted in statistical inference~\cite{gutman1989asymptotically,li2014,zhou2020second,zhou2022second} and the final decision step involves a potential exhaustive search step that can be further simplified. Secondly, when the number of outliers is unknown, the case of zero outlier was not considered in \cite{bu2019linear}, making the test design and theoretical analysis incomplete. We address both limitations in this paper. For the sequential setting, the low complexity test was not studied previously. We fill the research gap in this paper by proposing low complexity exponentially consistent sequential tests and analyzing their large deviations performance. Our results reveal the benefit of sequentiality by showing that our proposed low-complexity sequential tests achieve better performance than the low-complexity fixed-length tests for both cases of known and unknown number of outliers. Our main contributions are summarized with further details in the next section.

\subsection{Main Contributions}
In a nutshell, for outlier hypothesis testing, we propose low complexity non-parametric fixed-length and sequential tests for both cases of known and unknown number of outliers, show that our tests are exponentially consistent and demonstrate the superior performance of our tests in balancing detection performance and computational complexity. Our theoretical results reveal the benefit of sequentiality and the penalty of not knowing the number of outliers.

We first consider the case with known number of outliers. For the fixed-length setting, we strengthen the results of~\cite{bu2019linear} by allowing the test to use either KL divergence or GJS divergence and replacing the exhaustive search step of the test in \cite[Algorithm 2]{bu2019linear} with a sorting procedure. GJS divergence is widely adopted in non-parametric statistical inference~\cite{gutman1989asymptotically,li2014,li2017universal,zhou2020second,zhou2022second,zhou2024large} due to its connection to generalized likelihood ratio test while the sorting procedure ensures the same performance with much reduced complexity. In Fig. \ref{comp:li}, we numerically verify that our fixed-length test strikes a much better tradeoff between detection performance and computation complexity with respect to the optimal fixed-length test in~\cite[Eq. (37)]{li2014} that uses exhaustive search.  Furthermore, Fig. \ref{kl_vs_gjs} shows that using GJS divergence to construct scoring functions enables better detection performance in certain cases. For the sequential setting, we propose a novel non-parametric low-complexity test, show that our test has bounded expected stopping time for any pair of unknown nominal and anomalous distributions, and characterize the large deviations performance of our test. Our low-complexity sequential test strikes a better tradeoff between detection performance and computation complexity than the optimal sequential test in~\cite[Eq. (43)]{diao2025sequential}. Finally, comparing our results for low-complexity fixed-length and sequential tests, we analytically demonstrate the benefit of sequentiality and numerically illustrate the benefit in Figs. \ref{known_exponent} and \ref{low_compare}.

We next generalize the above results to the case of unknown number of outliers. In this case, there are three error events~\cite{zhou2020second}: misclassification, false reject and false alarm. A misclassification event occurs when the test identifies an incorrect set of outliers, a false reject event occurs when the test incorrectly claims no outlier while there exists outliers, and a false alarm event occurs when the test incorrectly claims existence of outliers while there is no outlier. For the fixed-length setting, we strengthened the result in~\cite{bu2019linear} by removing the implicit assumption of positive number of outliers, adding an outlier detection phase in the test design, and analyzing the large deviations performance of our tests to reveal the exponent tradeoff for probabilities of three error events. For the sequential setting, we propose a novel non-parametric test that has bounded expected stopping time under mild conditions, characterize the exponent tradeoff of three error probabilities and reveal the benefit of sequentiality (cf. Figs. \ref{unknown_exponent} and \ref{low_compare_unknown}). Specifically, our sequential test consists of an outlier detection phase and an outlier identification phase. In outlier detection, our test checks whether there exists outliers by comparing the maximal pairwise scoring function value with a positive threshold. In outlier identification, we replace the computationally complicated enumeration procedure in~\cite[Eq. (93)]{diao2025sequential} with a simpler procedure of comparing each pairwise scoring function with another two positive thresholds, which classifies each sequence as an outlier or a nominal sample. This way, our sequential test has polynomial complexity with respect to the total number of observed sequences regardless of the number of outliers and achieves a much better tradeoff between detection performance and computational complexity than the sequential test in~\cite[Eq. (93)]{diao2025sequential}. Finally, for both fixed-length and sequential tests, we theoretically reveal the penalty of not knowing the number of outliers on the detection performance by comparing our results with known and unknown number of outliers, and numerically illustrate the penalty in the second remark below Theorem \ref{fixed:unknown} and the first remark below Theorem \ref{th:unknown}, respectively.

\subsection{Other Related Studies}

We briefly recall other (non-exhausting) related studies on OHT. Zhang, Diao and Zhou~\cite{zhang2024large} studied the impact of distribution uncertainty on the large deviations performance of optimal fixed-length tests. Tajer, Veeravalli and Poor~\cite{tajer2014outlying} proposed a data-driven framework for OHT in large datasets and proposed adaptive and universal detection strategies. When the observed sequences are continuous, Zou~\emph{et al.}~\cite{zou2017nonparametric} proposed a non-parametric fixed-length test using the maximum mean discrepancy metric~\cite{gretton2012kernel}. Recently, Zhu and Zhou~\cite{zhu2024exponentially} refined the results in~\cite{zou2017nonparametric} by proposing a fixed-length test with better detection performance and proposing exponentially consistent two-phase~\cite{lalitha2016reliability,diao2023achievable,diao2023classification} and sequential tests.

OHT is also related with statistical classification. In particular, statistical classification, the non-parametric version of hypothesis testing, was initiated by Gutman~\cite{gutman1989asymptotically} who proposed a fixed-length test and proved its optimality in the generalized Neyman-Pearson sense. Zhou, Tan and Motani~\cite{zhou2020second} refined Gutman's result by deriving second-order asymptotic result that approximates the detection performance of optimal tests with finite sample sizes. The above results have been generalized to the case with distribution uncertainty~\cite{hsu2020binary} and sequential setting~\cite{mahdi2021sequential,Ihwang2022sequential,li2025unified}.

\subsection*{Notation}
We use $\bbR_+$ and $\bbN$ to denote the sets of non-negative real numbers and natural numbers, respectively. Given any two integers $(a,b)\in\bbN^2$ such that $1\leq a\leq b$, we use $[a:b]$ to denote the set of integers $\{a,a+1,\ldots,b\}$ and use $[a]$ to denote $[1:a]$. Random variables and their realizations are denoted by upper case variables (e.g., $X$) and lower case variables (e.g., $x$), respectively. All sets are denoted in calligraphic font (e.g., $\mathcal{X}$). Given any set $\calX$, we use $\calX^c$ to denote its complement. Given any integer $n\in\bbN$, let $X^n:=(X_1,\ldots X_n)$ be a random vector of length $n$ and let $x^n=(x_1,\ldots,x_n)$ be a particular realization. The set of all probability distributions on a finite set $\calX$ is denoted as $\calP(\calX)$. Given a sequence $x^n\in\calX^n$, the type or empirical distribution $\hatT_{x^n}$ is defined such that for each $a\in\calX$, $\hatT_{x^n}(a)=\frac{1}{n}\sum_{i=1}^{n}\mathbbm{1}(x_i=a)$. The set of types formed from length-$n$ sequences with alphabet $\calX$ is denoted by $\calP^n(\calX)$. Given any $P\in\calP^n(\calX)$, the set of all sequences of length $n$ with type $P$, a.k.a. the type class, is denoted by $\calT_P^n$.

\section{Problem Formulation and Existing Results}
\label{sec:pf}

Fix two integers $(n,M)\in\bbN^2$ and two distributions $(P_\rmN,P_\rmA)\in\calP(\calX)^2$. In outlier hypothesis testing, one is given a set of $M$ observed sequences $\bX^{\tau}:=\big\{X^{\tau}_1,\ldots,X^{\tau}_M\big\}$, where $\tau$ is a random stopping time with respect to the filtration $\{\sigma\{X_1,X_2,\ldots X_n\}\}_{n\in\bbN}$. The majority of the $M$ sequences are nominal samples generated i.i.d. from a nominal distribution $P_\rmN$ while the rest few outliers are generated i.i.d. from another anomalous distribution $P_\rmA$. The task of OHT is to design a non-parametric test to reliably identify all outliers or claim there is no outlier.

\subsection{Problem Formulation: Case of Known Number of Outliers}

Fix any integer $t\in\bbN$ such that $0<t\le\lceil\frac{M}{2}-1\rceil$. Assume that there are $t$ outliers among $M$ observed sequences. Let $\calS(t)$ denote the set of all subsets of $[M]$ with size $t$, i.e.,
\begin{align}\label{S_t}
\calS(t):=\{\calB\subset[M]:|\calB|=t\}.
\end{align}
Our task is to design a non-parametric test $\Phi:~\calX^{M\tau}\to\{\{\rmH_\calB\}_{\calB\in\calS(t)}\}$ to determine which sequences are outliers, where for each $\calB\in\calS(t)$, the hypothesis $\rmH_{\calB}$ means that for all $j\in\calB$, the $j$-th sequence is an outlier.

Fix any $\calB\in\calS(t)$. Define a set
\begin{align}\label{def:calM_calB}
\calM_\calB:=[M]\backslash\calB=\{j\in[M]:j\notin \calB\}.
\end{align}
To evaluate the performance of a test, we consider the following misclassification probability:
\begin{align}\label{known:beta}
\beta_\calB(\Phi|P_\rmN,P_\rmA):=\bbP_\calB\{\Phi(\bX^\tau)\neq\rmH_\calB\},
\end{align}
where we define $\bbP_\calB(\cdot):=\Pr\{\cdot|\rmH_\calB\}$ to denote the joint distribution of observed sequences $\bX^\tau$, where for each $i\in\calB$, $X_i^\tau$ is generated i.i.d. from the anomalous distribution $P_\rmA$ and for each $j\in\calM_\calB$, $X_j^\tau$ is generated i.i.d. from the nominal distribution $P_\rmN$. The misclassification probability $\beta_\calB(\cdot)$ is the probability that the test $\Phi$ fails to identify the true set of outliers. Furthermore, since the random stopping time could be rather large, we need to bound the following expected stopping time:
\begin{align}\label{def:EB}
\mathbb{E}_\calB[\tau]&=\sum_{k=1}^{\infty}\bbP_\calB\{\tau> k\}.
\end{align}
One would require the expected stopping time to be bounded so that the test stops in finite time on average. The following definition specifies such constraint.
\begin{definition}
A sequential test $\Phi$ is said to satisfy the expected stopping time universality constraint if there exists an integer $n\in\bbN$ such that for any pair of distributions $(P_\rmN,P_\rmA)\in\calP(\calX)^2$,
\begin{align}\label{constraint2:est}
\max_{\calB\in\calS(t)}\mathbb{E}_\calB[\tau]\le n.
\end{align}
\end{definition}

For a sequential test satisfying the expected stopping time universality constraint, the theoretical benchmark is the following misclassification exponent that characterizes the exponential decay rate of the misclassification probability:
\begin{align}
E_\calB(\Phi|P_\rmN,P_\rmA):=\liminf_{n\to\infty}\frac{-\log\beta_\calB(\Phi|P_\rmN,P_\rmA)}{n}.
\end{align}
When $\tau=n$ is fixed a priori for some integer $n\in\bbN$, the test reduces to a fixed-length test, which naturally satisfies the expected stopping time universality constraint. In this paper, we study both fixed-length and sequential tests.

\subsection{Problem Formulation: Case of Unknown Number of Outliers}
Fix an integer $T\in\bbN$ such that $0<T\le\lceil\frac{M}{2}-1\rceil$. Assume that there are at most $T$ outliers, i.e., the number of outliers is unknown but upper bounded by $T$. Recall the definitions of the set $\calS(t)$ in \eqref{S_t}. Define the union of sets $\calS(t)$ over $t\in[T]$ as
\begin{align}\label{S}
\calS:=\bigcup_{t\in[T]}\calS(t).
\end{align}
When the number of outliers is unknown, our task is to design a non-parametric test to identify the potential set of outliers and avoid false alarm. In other words, we need to design a test $\Phi:~\calX^{M\tau}\to\{\{\rmH_\calB\}_{\calB\in\calS},\mathrm{H_r}\}$ to
classify among the following $|\calS|+1$ hypotheses:
\begin{itemize}
\item $\rmH_{\calB}$,~$\calB\in\calS$: for each $j\in\calB$, the $j$-th sequence is an outlier.
\item $\mathrm{H_r}$: there is no outlier.
\end{itemize}

To evaluate the performance of a test, for each $\calB\in\calS$, we consider the following misclassification and false reject probabilities under the non-null hypothesis $\rmH_{\calB}$:
\begin{align}
\beta_\calB(\Phi|P_\rmN,P_\rmA)&:=\bbP_\calB\{\Phi(\bX^\tau)\notin\{\rmH_\calB,\rmH_\rmr\}\},\\
\zeta_\calB(\Phi|P_\rmN,P_\rmA)&:=\bbP_\calB\{\Phi(\bX^\tau)=\rmH_\rmr\},
\end{align}
where $\bbP_\calB(\cdot)$ is defined similarly as in \eqref{known:beta}.
The misclassification probability $\beta_\calB(\cdot)$ bounds the probability that the test $\Phi$ identifies an incorrect set of outliers while the false reject probability $\zeta_\calB(\cdot)$ bounds the probability that the test $\Phi$ falsely claims there is no outlier. Under the null hypothesis, we have the false alarm probability:
\begin{align}
\rmP_{\rm{fa}}(\Phi|P_\rmN,P_\rmA)&:=\bbP_\rmr\{\Phi(\bX^\tau)\ne\rmH_\rmr\},
\end{align}
where we define $\bbP_\rmr(\cdot):=\Pr\{\cdot|\mathrm{H_r}\}$ to denote the joint distribution of observed sequences $\bX^\tau$, where for all $j\in[M]$, $X_j^\tau$ is generated i.i.d. from the nominal distribution $P_\rmN$. The false alarm probability $\rmP_{\rm{fa}}(\cdot)$ bounds the probability that the test $\Phi$ falsely claims the existence of outliers while there is no outlier.

Furthermore, we also need to control the following expected stopping times under each non-null hypothesis $\rmH_\calB$ and the null hypothesis $\mathrm{H_r}$:
\begin{align}
\mathbb{E}_\calB[\tau]&=\sum_{k=1}^{\infty}\bbP_\calB\{\tau> k\},\\
\mathbb{E}_\rmr[\tau]&=\sum_{k=1}^{\infty}\bbP_\rmr\{\tau> k\}.
\end{align}
The constraint is specified in the following definition.
\begin{definition}
A sequential test $\Phi$ is said to satisfy the expected stopping time universality constraint if there exists an integer $n\in\bbN$ such that for any pair of distributions $(P_\rmN,P_\rmA)\in\calP(\calX)^2$,
\begin{align}
\max\Big\{\max_{\calB\in\calS} \mathbb{E}_\calB[\tau],\mathbb{E}_\rmr[\tau]\Big\}\le n.
\end{align}
\end{definition}

For a sequential test satisfying the expected stopping time universality constraint, the theoretical benchmarks are the following error exponents that characterize the exponential decay rates for the probabilities of misclassification, false reject and false alarm:
\begin{align}
E_{\beta_\calB}(\Phi|P_\rmN,P_\rmA)&:=\liminf_{n\to\infty}\frac{-\log\beta_\calB(\Phi|P_\rmN,P_\rmA)}{n},~\calB\in\calS,\label{def:mis}\\
E_{\zeta_\calB}(\Phi|P_\rmN,P_\rmA)&:=\liminf_{n\to\infty}\frac{-\log\zeta_\calB(\Phi|P_\rmN,P_\rmA)}{n},~\calB\in\calS,\label{def:fr}\\
E_{\rm fa}(\Phi|P_\rmN,P_\rmA)&:=\liminf_{n\to\infty}\frac{-\log\mathrm{P_{fa}}(\Phi|P_\rmN,P_\rmA)}{n}\label{def:fa}.
\end{align}
Similarly to the case of known number of outliers, when $\tau=n$ is fixed a priori for some integer $n\in\bbN$, the test reduces to a fixed-length test.

\subsection{Existing Fixed-length Tests}
In this section, we recall two existing fixed-length tests~\cite{li2014,zhou2022second} that are proved optimal under certain conditions.

When the number of outliers is known, Li, Nitinawarat and Veeravalli~\cite[Eq. (37)]{li2014} proposed a fixed-length test and proved its optimality when the total number $M$ of observed sequences tends to infinity. Recall the definitions of $\calM_\calB=\{i\in[M]:i\notin\calB\}$ and $\calS(t)=\{\calB\subset[M]:~|\calB|=t\}$. Given a tuple of distributions $\bQ=(Q_1,\ldots,Q_M)\in\calP(\calX)^M$, for each $\calB\in\calS(t)$, define a scoring function
\begin{align}
\rmG_{\mathrm{Li},\calB}(\bQ):=\sum_{j\in\calM_\calB}D\bigg(Q_j\Big\|\frac{\sum_{l\in\calM_\calB}Q_l}{M-|\calB|}\bigg)\label{GliB}.
\end{align}
Note that $\rmG_{\mathrm{Li},\calB}(\bQ)$ measures the similarity of distributions $\{Q_i\}_{i\in\calM_\calB}$, which equals zero if and only if $Q_j=Q$ for all $j\in\calM_\calB$ with an arbitrary $Q\in\calP(\calX)$. Using types of observed sequences $\bx^n=(x_1^n,\ldots,x_M^n)$, Li, Nitinawarat and Veeravalli~\cite[Eq. (37)]{li2014} proposed the following fixed-length test using the minimal scoring function decision rule:
\begin{align}\label{test_Li}
\Phi_{\rm Li}(\bx^n)=\rmH_\calC,~\mathrm{if}~\calC=\argmin\limits_{\calB\in\calS(t)}\rmG_{\mathrm{Li},\calB}\big(\hatT_{x_1^k},\ldots,\hatT_{x_M^k}\big).
\end{align}

When the number of outliers is unknown, Zhou, Wei and Hero~\cite[Eq. (43)]{zhou2022second} proposed an optimal fixed-length test in the generalized Neyman-Pearson sense. Given a tuple of distributions $\bQ=(Q_1,\ldots,Q_M)\in\calP(\calX)^M$, for each $\calB\in\calS$, define another scoring function
\begin{align}\label{GB}
\rmG_\calB(\bQ):=\sum_{i\in\calB}D\bigg(Q_i\Big\|\frac{\sum_{t\in\calB}Q_t}{|\calB|}\bigg)+\sum_{j\in\calM_\calB}D\bigg(Q_j\Big\|\frac{\sum_{l\in\calM_\calB}Q_l}{M-|\calB|}\bigg).
\end{align}
Analogously to $\rmG_{\mathrm{Li},\calB}$, $\rmG_\calB(\bQ)$ measures the similarity of distributions $\{Q_i\}_{i\in\calB}$ and $\{Q_j\}_{j\in\calM_\calB}$, which equals zero if and only if $Q_j=Q_1$ for all $j\in\calM_\calB$ and $Q_i=Q_2$ for all $i\in\calB$ with arbitrary distributions $(Q_1,Q_2)\in\calP(\calX)^2$. Using types of observed sequences $\bx^n=(x_1^n,\ldots,x_M^n)$ and a positive real number $\lambda\in\bbR_+$, Zhou, Wei and Hero proposed the following fixed-length test $\Phi_{\mathrm{Zhou}}$~\cite[Eq. (43)]{zhou2022second}:
\begin{align}
\Phi_{\mathrm{Zhou}}({\bx^n}):=\left\{
\begin{array}{lll}
\rmH_\calB,&\mathrm{if}~\rmS_\calB(\bx^n)<\mathop{\min}\limits_{\calC\in\calS_\calB}\rmS_{\calC}({\bx^n})~\mathrm{and}\mathop{\min}\limits_{\calC\in \calS_\calB} \rmS_{\calC}({\bx^n})>\lambda,\\
\rmH_\rmr, &\rm{otherwise},
\end{array}
\right.\label{test_zhou}
\end{align}
where $\calS_\calB:=\calS\setminus\{\calB\}=\{\calC\in\calS:\calC\ne\calB\}$ and the scoring function $\rmS_\calB({\bX^n})$ is defined as
\begin{align}
\label{s_b}
\rmS_\calB({\bX^n}):=\mathrm{G}_{\calB}(\hat{T}_{x_1^n},\ldots,\hat{T}_{x^n_M}).
\end{align}

\subsection{Existing Sequential Tests}
\label{sec:existing_seq}
When the number of outliers is known as $t$, Diao and Zhou~\cite[Eq. (41) and (43)]{diao2025sequential} proposed an optimal sequential test satisfying the expected stopping time universality constraint. The sequential test $\Phi_\mathrm{Diao}=(\tau,\phi)$ consists of a random stopping time and decision rule. The stopping time $\tau$ is defined as
\begin{align}\label{tau:optimal}
\tau:=\inf\{k\ge n-1: \exists~\calC\in\calS(t)~\mathrm{s.t.}~\rmS_\calC(\bx^k)\le f(k)\}.
\end{align}
where $f(k):=\frac{(M+1)|\calX|\log (k+1)}{k}$. At the stopping time $\tau$, the test applies the minimal decision rule as follows:
\begin{align}\label{test:optimal}
\phi(\bx^\tau)=\rmH_\calB,~\mathrm{if}~\calB=\argmin\limits_{\calC\in\calS(t)}\rmS_\calC(\bx^\tau).
\end{align}

When the number of outliers is unknown but an upper bound $T$ is known, Diao and Zhou~\cite[Eq. (92) and (93)]{diao2025sequential} proposed the following sequential test $\Phi_\mathrm{Diao}^\mathrm{u}$ satisfying the expected stopping time universality constraint.
Given two positive real numbers $(\lambda_1,\lambda_2)\in\bbR_+^2$ such that $\lambda_1\leq \lambda_2$, the stopping time $\tau$ is defined as follows:
\begin{align}\label{tau:unknown}
\tau:=\inf\Big\{k\ge n-1: \exists~\calC\in\calS~\mathrm{s.t.}~\rmS_\calC(\bx^k)\le\lambda_1~\mathrm{and}~\min_{\calD\in\calS_\calC}\rmS_\calC(\bx^k)>\lambda_2,~\mathrm{or}~\forall~\calC\in\calS~\mathrm{s.t.}~\rmS_\calC(\bx^k)\le\lambda_1\Big\}.
\end{align}
At the stopping time $\tau$, the test uses the following decision rule:
\begin{align}\label{test:unknown}
\phi(\bx^\tau)=\left\{
\begin{array}{cl}
\rmH_\calB&\mathrm{if}~\rmS_\calB(\bx^k)\le\lambda_1,~\mathrm{and}~\min_{\calC\in\calS_\calB}\rmS_\calC(\bx^k)>\lambda_2,\\
\rmH_\rmr&\mathrm{Otherwise}.
\end{array}
\right.
\end{align}

Although the above fixed-length and sequential tests are all exponentially consistent and optimality guarantees are provided when the number of outliers is known, these tests suffer from prohibitively high computational complexity due to the use of exhaustive search. Specifically, there are $M\choose t$ possibilities when the number of outliers is known and $\sum_{i=1}^{T}{M\choose i}$ possibilities when the number of outliers is unknown. For example, when $M=100$, when it is known that there are  $t=10$ outliers, ${M\choose t}=1.731\times 10^{13}$; when an upper bound $T=20$ is known, $\sum_{i=1}^{T}{M\choose i}=1.347\times 10^{29}$. With a further step towards practical applications, to address the above problem, we propose low complexity exponentially consistent tests.

\section{Main Results for the Case of Known Number of Outliers}
\subsection{Preliminaries}

Fix any pair of distributions $(P,Q)\in\calP(\calX)^2$. Let $f:~\calP(\calX)^2\to\bbR_+$ be a scoring function such that $f(P,Q)=0$ if and only if $P=Q$ and $f(P,Q)>0$ if $P\neq Q$. Such function includes Kullback-Leibler (KL) divergence~\cite{bu2019linear} and generalized Jensen-Shannon (GJS) divergence~\cite{gutman1989asymptotically,zhou2020second}.
\begin{enumerate}
\item The KL divergence is defined as
\begin{align}\label{KL}
D(P\|Q):=\sum_{x\in\calX}P(x)\log\frac{P(x)}{Q(x)}.
\end{align}
KL divergence is extensively used for parametric statistical inference problems including hypothesis testing~\cite{tuncel2005error,lalitha2016reliability,liao2017hypothesisMI,grigoryan2011multiple}.
\item
The GJS divergence~\cite[Eq. (2.3)]{zhou2020second} is defined as
\begin{align} \label{GJS}
{\rm GJS}(P,Q,1):=D\left(P\Big\| \frac{P+Q}{2}\right)+D\left(Q\Big \| \frac{P+Q}{2}\right),
\end{align}
which also has the following variation form~\cite[Eq. (6)]{Ihwang2022sequential}
\begin{align}\label{variation}
{\rm GJS}(P,Q,1)=\min_{V\in\calP(\calX)} D(P||V)+D(Q||V).
\end{align}
GJS divergence is widely used for non-parametric statistical inference problems including  classification~\cite{gutman1989asymptotically,zhou2020second,bai2022achievable,diao2023classification} and sequence matching~\cite{zhou2024large,unnikrishnan2015asymptotically}. Note that ${\rm GJS}(P,Q,1)$ is symmetric while $D(Q||P)$ is not.
\end{enumerate}
In this paper, we consider both measures in our theoretical analyses.

\subsection{Low Complexity Fixed-length Test}\label{sec:known_fix}

\subsubsection{Test Design and Asymptotic Intuition}
Recall that $M$ is the total number of observed sequences, $t$ is the number of outliers, and the set $\calS(t)$ was defined in \eqref{S_t}. The fixed-length test in Algorithm \ref{FA:known} is essentially the test in~\cite[Algorithm 2]{bu2019linear} except that i) we generalize the scoring function from KL divergence to other functions including the GJS divergence and ii) we replace the step of exhaustive search over all sets $\calS(t)$ with an equivalent but simpler step of finding the smallest $t$ elements from a set of size $M$. As we shall show in Fig. \ref{kl_vs_gjs}, using GJS divergence as the scoring function can yield better performance in certain cases. We would like to emphasize that our main contributions in this paper lie in the study of sequential tests and the fixed-length test with unknown number of outliers. The fixed-length test for the case of known number of outliers serves as the benchmark and is included for the completeness of the story so that we can reveal the benefit of sequentiality and the penalty of not knowing the number of outliers.

The key steps of the test are summarized as follows. Recall that $[M]=\{1,\ldots,M\}$. In steps 1-4 of Algorithm \ref{FA:known},  with high probability, the test chooses a nominal sample $x_{i^*}^n$ that is generated i.i.d. from the unknown nominal distribution $P_\rmN$, as we shall explain shortly. Subsequently, in steps 5-6, the test calculates $M$
scoring function values and outputs the indices of the $t$ sequences that have $t$ largest scoring function values.

\begin{algorithm}[tb]
\caption{Low complexity fixed-length test $\Phi_{\mathrm{fix}}$ with known number of outliers}
\label{FA:known}
\begin{algorithmic}[1]
\Require $M$ observed sequences $(x_1^n,\ldots,x_M^n)$ and the number $t$ of outliers
\Ensure The set $\calB\in\calS(t)$ of indices for outliers.
\State Choose a number $l\in[M]$ randomly and set $\hatT_0=\hatT_{x_l^n}$
\State Compute $\{f\big(\hatT_{x_i^n},\hatT_0\big)\}_{i\in[M]}$ and sort the values in a non-increasing order to form the vector $\bv_1$
\State Set $i^*$ as the index of the sequence corresponding to the $\lceil\frac{M}{2}\rceil$-th element of $\bv_1$
\State Set $\tilP_\rmN=\hatT_{x_{i^*}^n}$
\State Compute $\{f\big(\hatT_{x_i^n},\tilP_\rmN\big)\}_{i\in[M]}$ and sort the values in a non-increasing order to form another vector $\bv_2$
\State Set $\calB$ as the set that includes indices of sequences corresponding to the first $t$ elements of $\bv_2$
\end{algorithmic}
\end{algorithm}

We now explain why the above test works asymptotically using the weak law of large numbers. Fix any set $\calB\in\calS(t)$ and recall the definition of $\calM_\calB$ was defined in \eqref{def:calM_calB}. As the sample size $n$ increases, under hypothesis $\rmH_\calB$, for each $i\in\calB$, the type $\hat{T}_{X_i^n}$ of the outlier $X_i^n$ converges in probability to the unknown anomalous distribution $P_\rmA$, while for each $j\in\calM_\calB$ , the type $\hat{T}_{X_j^n}$ of the nominal sample $x_j^n$ converges in probability to the unknown nominal distribution $P_\rmN$. Thus, for any $(i,j)\in\calB^2$ or any $(i,j)\in\calM_\calB^2$ such that $i\neq j$, the scoring function $f(\hatT_{X_i^n},\hatT_{X_j^n})$ converges to zero while for any $(i,j)\in\calB\times\calM_\calB$, the scoring function $f(\hatT_{X_i^n},\hatT_{X_j^n})$ converges to a positive real number. Therefore, considering the fact that there are $t<\frac{M}{2}$ outliers among $M$ observed sequences, asymptotically with probability one, the distribution $\tilP_N$ chosen in step 4 of Algorithm \ref{FA:known} is the type of a nominal sample and the set $\calB$ collects all outliers.

\subsubsection{Theoretical Results and Discussions}

Fix any pair of distributions $(P_1,P_2)\in\calP(\calX)^2$. Define the following exponent function
\begin{align}\label{eta0}
\eta(P_1,P_2):=\min_{\substack{(Q_1,Q_2,Q_3)\in\calP(\calX)^3:f(Q_1,Q_2)\le f(Q_3,Q_2)}}D(Q_1||P_1)+D(Q_2||P_2)+D(Q_3||P_2).
\end{align}
Note that $\eta(P_1,P_2)$ is strictly positive when $P_1\neq P_2$.

\begin{theorem}\label{fixed:known}
Under any pair of nominal and anomalous distributions $(P_\rmN,P_\rmA)\in\calP(\calX)^2$, for any $\calB\in\calS(t)$, the misclassification exponent of the fixed-length test in Algorithm \ref{FA:known} satisfies
\begin{align}
E_\calB\big(\Phi_{\mathrm{fix}}|P_\rmN,P_\rmA\big)&\ge\min\big\{\eta(P_\rmA,P_\rmN),~\eta(P_\rmN,P_\rmA)\big\}.
\end{align}
\end{theorem}
The proof of Theorem \ref{fixed:known} is similar to \cite[Appendix B]{bu2019linear} and provided in Appendix \ref{proof:fix_known} for completeness. When the scoring function $f(\cdot)$ is the KL divergence, Theorem \ref{fixed:known} is exactly the achievability part of \cite[Theorem 1]{bu2019linear}.

Theorem \ref{fixed:known} shows that the misclassification exponent of the low-complexity test in Algorithm \ref{FA:known} is lower bounded by the minimization of two exponent functions: $\eta(P_\rmN,P_\rmA)$ and $\eta(P_\rmA,P_\rmN)$. We next explain why these two exponent functions appear. Given the test in Algorithm \ref{FA:known}, there are two error events: $\calE^\mathrm{f,k}_1$ where in step 4, the test chooses $\tilP_\rmN$ as the type of an outlier, and $\calE^\mathrm{f,k}_2$ where in step 7, the test classifies a nominal sample as an outlier. The error event $\calE^\mathrm{f,k}_1$ can be further categorized into two events: $\calE^\mathrm{f,k}_{1,1}$ when $\hatT_0$ in step 1 is the type of a nominal sample and $\calE^\mathrm{f,k}_{1,2}$ when $\hatT_0$ in step 1 is the type of an outlier. The exponential decay rates for the probabilities of error events $\calE^\mathrm{f,k}_{1,1}$ and $\calE^\mathrm{f,k}_{1,2}$ are lower bounded by $\eta(P_\rmN,P_\rmA)$ and $\eta(P_\rmA,P_\rmN)$, respectively. Analogously, the exponential decay rate for the probability of error event $\calE^\mathrm{f,k}_2$ is lower bounded by $\eta(P_\rmN,P_\rmA)$.

\begin{figure}[tb]
\centering
\includegraphics[width=.5\linewidth]{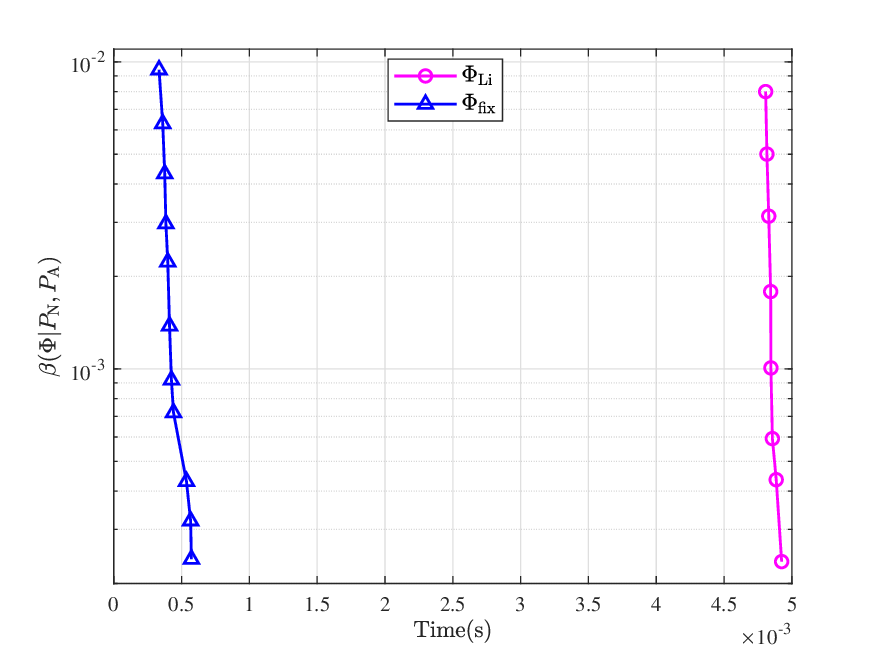}
\caption{Plot of the simulated misclassification probabilities as a function of running times of the fixed-length test in Algorithm \ref{FA:known} and the fixed-length test $\Phi_{\rm Li}$ in \eqref{test_Li} when $M=10$ and $t=3$ and $(P_\rmN,P_\rmA)=\mathrm{Bern}(0.23,0.3)$. As observed, the low-complexity test in Algorithm \ref{FA:known} achieves the same misclassification probability with much less running time than the test $\Phi_{\rm Li}$.}
\label{comp:li}
\end{figure}

The low-complexity test in Algorithm \ref{FA:known} has smaller computational complexity than the existing fixed-length test $\Phi_{\rm Li}$ in \eqref{test_Li}. In particular, the test $\Phi_{\rm Li}$ applies exhaustive search to identify the set of outliers, whose computational complexity is proportional to $M\choose t$. In contrast, the test $\Phi_{\mathrm{fix}}$ in Algorithm \ref{FA:known} has polynomial complexity in $M$, which is highly practical. To illustrate, in Fig. \ref{comp:li}, we plot the simulated misclassification probabilities and running times for the test in Algorithm \ref{FA:known} and the test $\Phi_{\rm Li}$ in \eqref{test_Li} when $(P_\rmN,P_\rmA)=\mathrm{Bern}(0.23,0.3)$, our test use GJS divergence as the scoring function where $\Phi_{\rm Li}$ uses a similar scoring function (cf. \eqref{GliB}). As observed, the low-complexity test in Algorithm \ref{FA:known} achieves a much better tradeoff between misclassification probability and running time than the test $\Phi_{\rm Li}$ in \eqref{test_Li}.

\begin{figure}[tb]
\centering
\includegraphics[width=.5\linewidth]{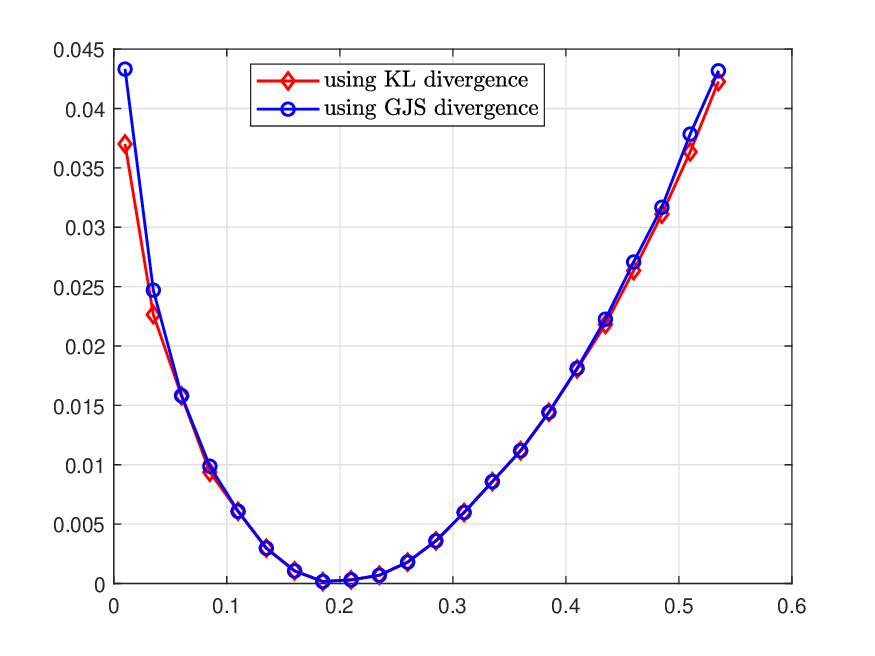}
\caption{Numerical comparison of achievable misclassification exponents in Theorem \ref{fixed:known} for KL and GJS divergence scoring functions when $P_\rmN=\rm{Bern}(0.2)$ and $P_\rmA=\rm{Bern}(a)$ for different values of $a\in[0.01,0.55]$ such that $a\neq 0.2$. As observed, GJS divergence scoring function can yield larger misclassification exponent in certain cases.}
\label{kl_vs_gjs}
\end{figure}

Finally, we numerically compare the achievable misclassification exponents when different scoring functions are used. Specifically, in Fig. \ref{kl_vs_gjs}, the exponents in Theorem \ref{fixed:known} are calculated for KL and GJS divergence scoring functions when the nominal distribution is $P_\rmN=\rm{Bern}(0.2)$ and the anomalous distribution is $P_\rmA=\rm{Bern}(a)$, where $a\in[0.01,0.55]$ and $a\neq 0.2$. As observed, the misclassification exponents depend on the unknown generating distributions and GJS divergence scoring function can yield better performance in certain cases. In fact, GJS divergence is extensively used to construct optimal tests for statistical classification~\cite{gutman1989asymptotically,zhou2020second,Ihwang2022sequential} and the low-complexity test in Algorithm \ref{FA:known} is closely related to statistical classification.

\subsection{Low Complexity Sequential Test}\label{sec:known_sequential}

\subsubsection{Test Design and Asymptotic Intuition}
This subsection presents our low-complexity sequential test that satisfies the expected stopping time universality constraint. Given parameters $(\lambda_1,\lambda_2,n)\in\bbR_+^2\times\bbN$ such that $\lambda_1\leq \lambda_2$, our sequential test $\Phi_\mathrm{seq}=(\tau,\phi)$ is summarized in Algorithm \ref{CA:known}. Consistent with sequential test design for statistical classification~\cite{Ihwang2022sequential}, we set the initial sample size as $k=n-1$ to avoid early stopping. Subsequently, our test randomly chooses a sequence, whose type is denoted as $\hatT_0$, and calculates $M$ scoring function values using the type of each observed sequence and $\hatT_0$. Subsequently, in steps 6-13, our test classifies each sequence as either a nominal sample or an outlier via two sets $(\calC_1,\calC_2)$ using binary classification with thresholds $(\lambda,\lambda_2)$. Our test stops if both sets $\calC_1$ and $\calC_2$ contain at least $t$ elements; otherwise, our test collects additional symbols and iterates from step 3. When our step stops, in steps 21-27, the final decision is made by outputting the indices of sequences that have $t$ largest or smallest scoring function values. Note that the sorting order differs since $\hatT_0$ chosen in step 4 can be the type of either an outlier or a nominal sample and we should account for both possibilities.

Our test in Algorithm \ref{CA:known} has much lower computational complexity than the optimal test $\Phi_{\rm Diao}$ in~\eqref{test:optimal} that uses exhaustive search. Specifically, our test in Algorithm \ref{CA:known} has polynomial complexity with respect to the total number $M$ of observed sequences while the optimal test in~\eqref{test:optimal} has complexity $M\choose t$.

\begin{algorithm}[tb]
\caption{Low complexity sequential test $\Phi_{\rm{seq}}$ with known number of outliers}
\label{CA:known}
\begin{algorithmic}[1]
\Require $M$ observed sequences, the number $t$ of outliers and parameters $(\lambda_1,\lambda_2,n)\in\bbR_+^2\times\bbN$
\Ensure The stopping time $\tau$ and the set $\calB$ for indices of outliers
\State Set $k=n-1$ and $\mathrm{flag}=0$
\State Collect observed sequences $(x_1^k,\ldots,x_M^k)$.
\While{$\mathrm{flag}=0$}
\State Choose a number $l\in[M]$ randomly and set $\hatT_0=\hatT_{x_l^k}$
\State Set $\calC_1=\emptyset$ and $\calC_2=\emptyset$
\For {$i\in[M]$}
\State Compute $f\big(\hatT_{x_i^k},\hatT_0\big)$
\If {$f\big(\hatT_{x_i^k},\hatT_0\big)\le\lambda_1$}
\State$\calC_1\leftarrow \calC_1\cup\{i\}$
\ElsIf {$f\big(\hatT_{x_i^k},\hatT_0\big)>\lambda_2$}
\State $\calC_2\leftarrow \calC_2\cup\{i\}$
\EndIf
\EndFor
\If {$\min\{|\calC_1|,|\calC_2|\}\ge t$}
\State $\mathrm{flag}=1$
\State break
\EndIf
\State Collect new symbols $(x_{1,k+1},\ldots,x_{M,k+1})$
\State Update $k$ as $k+1$
\EndWhile
\If {$|\calC_2|\ge|\calC_1|$}
\State Sort $\{f\big(\hatT_{x_i^k},\hatT_0\big)\}_{i\in\calC_1}$ in a non-decreasing order to form a vector $\bv$
\Else
\State Sort $\{f\big(\hatT_{x_i^k},\hatT_0\big)\}_{i\in\calC_2}$ in a non-increasing order to form a vector $\bv$
\EndIf
\State Set $\calC_\mathrm{out}$ as the set that includes indices of sequences corresponding to the first $t$ elements of $\bv$
\State \Return $\tau=k$ and $\calB=\calC_\mathrm{out}$
\end{algorithmic}
\end{algorithm}

We now explain why our test works asymptotically using the weak law of large numbers. As discussed in Sec. \ref{sec:known_fix}, as the sample size increases, for any two outliers or any two nominal samples, the scoring function converges to zero, which is less than any positive real number $\lambda_1$; otherwise, the scoring functions converge to a positive real number, which is greater than any $\lambda_2<\min\{f(P_\rmA,P_\rmN),~f(P_\rmN,P_\rmA)\}$. Thus, the correct set of outliers can be identified correctly.

\subsubsection{Theoretical Results and Discussions}

Fix any pair of distributions $(P_1,P_2)\in\calP(\calX)^2$. Given $\lambda\in\bbR_+$, define the following exponent function:
\begin{align}
\Omega(P_1,P_2,\lambda)&:=\min_{\substack{(Q_1,Q_2)\in\calP(\calX)^2:~f(Q_1,Q_2)\le\lambda}}D(Q_1\|P_1)+D(Q_2\|P_2).\label{omega}
\end{align}
The function $\Omega(P_1,P_2,\lambda)$ is non-increasing in $\lambda$. Specifically, $\Omega(P_1,P_2,\lambda)=0$ when $\lambda\geq f(P_1,P_2)$ while $\Omega(P_1,P_2,\lambda)$ achieves the following maximum value when $\lambda=0$, which is the R\'{e}nyi Divergence of order $\frac{\alpha}{1+\alpha}$~\cite[Eq. (7)]{Ihwang2022sequential}:
\begin{align}\label{maxomega}
\Omega(P_1,P_2,0)&=\min\limits_{\substack{Q\in\calP(\calX)}}D(Q||P_1)+D(Q||P_2)\\
&=D_{\frac{\alpha}{1+\alpha}}(P_1||P_2).
\end{align}

Furthermore, fix any distribution $P\in\calP(\calX)$. Given $\lambda\in\bbR_+$, define another exponent function:
\begin{align}
\Upsilon(P,\lambda)&:=\min_{\substack{(Q_1,Q_2)\in\calP(\calX)^2:~f(Q_1,Q_2)\ge\lambda}}D(Q_1\|P)+D(Q_2\|P).\label{upsilon}
\end{align}
The function $\Upsilon(P,\lambda)$ is non-decreasing in $\lambda$. In particular, $\Upsilon(P,\lambda)=0$ when $\lambda=0$ and $\Upsilon(P,\lambda)$ achieves the maximal value when $\lambda$ tends to infinity. When $f(P,Q)={\rm GJS}(P,Q,1)$, it follows from the variational formula of GJS divergence in \eqref{variation} that $\mathrm{GJS}(Q_1,Q_2,1)=\min_{V\in\calP(\calX)}D(Q_1||V)+D(Q_2||V)$ and thus,
\begin{align}
\Upsilon(P,\lambda)&=\min\limits_{\substack{(Q_1,Q_2)\in\calP(\calX)^2:\\
\mathrm{GJS}(Q_1,Q_2,1)\ge\lambda}}D(Q_1||P)+D(Q_2||P)\\
&\ge\lambda.
\end{align}

\begin{theorem}\label{th:known}
Under any pair of distributions $(P_\rmN,P_\rmA)\in\calP(\calX)^2$, given any pair of thresholds $(\lambda_1,\lambda_2)\in\bbR_+^2$ such that $\lambda_1\leq\lambda_2<\min\{f(P_\rmA,P_\rmN),~f(P_\rmN,P_\rmA)\}$, our low complexity sequential test in Algorithm \ref{CA:known} satisfies the expected stopping time universality constraint and ensures that for each $\calB\in\calS(t)$, the misclassification exponent satisfies
\begin{align}
E_\calB(\Phi_\mathrm{seq}|P_\rmN,P_\rmA)\ge\min\big\{\Omega(P_\rmN,P_\rmA,\lambda_1),~\Upsilon(P_\rmN,\lambda_2)\}.
\end{align}
\end{theorem}
The proof of Theorem \ref{th:known} is provided in Appendix \ref{proof:known}, where we extensively use the method of types~\cite{csiszar1998mt} to bound the expected stopping time and the exponential decay rate of misclassification probability of the test in Algorithm \ref{CA:known}. Theorem \ref{th:known} shows that the misclassification exponent of our sequential test in Algorithm \ref{CA:known} is lower bounded by the minimization of two exponent functions: $\Omega(P_\rmA,P_\rmN,\lambda_1)$ and $\Upsilon(P_\rmN,\lambda_2)$. This results from the analysis of the exponential decay rates for the following error event: $\calE^\mathrm{s,k}$ where in steps 8-12, our test claims a nominal sample as an outlier. In particular, $\Omega(P_\rmA,P_\rmN,\lambda_1)$ bounds the exponential decay rates for the probability of the error event $\calE^\mathrm{s,k}$ when $\hatT_0$ chosen randomly in step 4 is the type of an outlier, and $\Upsilon(P_\rmN,\lambda_2)$ bounds the exponential decay rates for the probability of the error event $\calE^\mathrm{s,k}$ when $\hatT_0$ is the type of a nominal sample.

\begin{figure}[tb]
\centering
\includegraphics[width=0.5\linewidth]{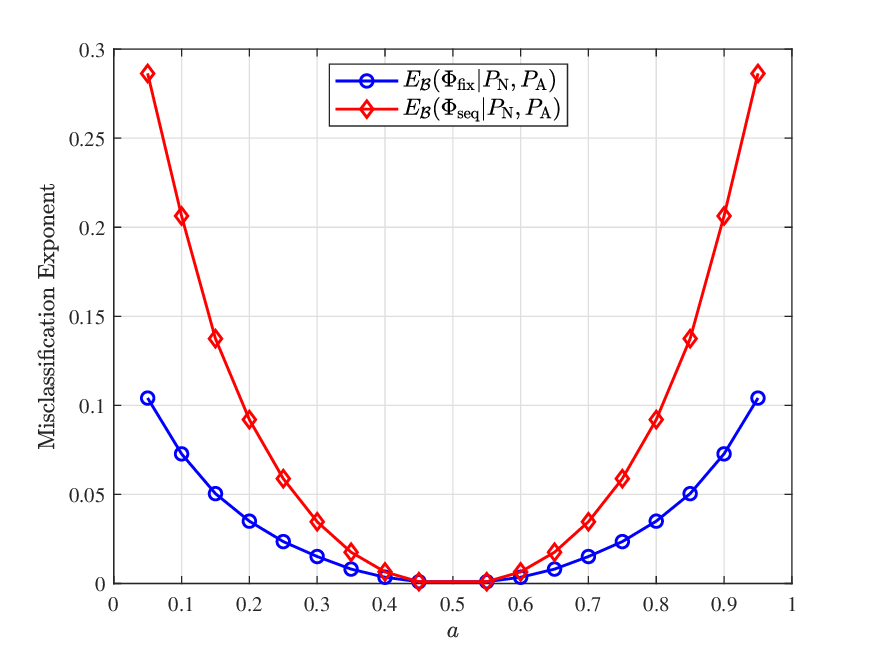}
\caption{Plot of achievable misclassification exponents for the sequential test in Theorem \ref{th:known} and the fixed-length test in Theorem \ref{fixed:known} when the scoring function $f(\cdot)$ is the GJS divergence, $P_\rmN=\rm{Bern}(0.5)$, $P_\rmA=\rm{Bern}(a)$ for $a\in(0,1)$ such that $a\neq 0.5$, $\lambda_1=0.0005$ and $\lambda_2=f(P_\rmA,P_\rmN)-0.0001$ for each $a$. As observed, the achievable misclassification exponent for the sequential test is larger than that for the fixed-length test.}
\label{known_exponent}
\end{figure}

We make several remarks. Firstly, the misclassification exponent in Theorem \ref{th:known} is maximized when $\lambda_1\to 0$ and $\lambda_2\to f(P_\rmA,P_\rmN)$ since $\Omega(P_\rmA,P_\rmN,\lambda)$ is non-increasing in $\lambda$ and $\Upsilon(P_\rmN,\lambda)$ is non-decreasing in $\lambda$. In particular, $\Omega(P_\rmA,P_\rmN,\lambda)$ achieves the maximal value $\Omega(P_\rmA,P_\rmN,0)$ in \eqref{maxomega} when $\lambda\to 0$ and $\Upsilon(P_\rmN,\lambda)$ achieves the maximal value $\Upsilon(P_\rmN,f(P_\rmA,P_\rmN))$ when $\lambda\to f(P_\rmA,P_\rmN)$. Thus, the maximal achievable misclassification exponent of our sequential test is
\begin{align}
\min\{\Omega(P_\rmA,P_\rmN,0),~\Upsilon(P_\rmN,f(P_\rmA,P_\rmN))\}\label{maxupsilon},
\end{align}
which is greater than the misclassification exponent $\min\big\{\eta(P_\rmA,P_\rmN),~\eta(P_\rmN,P_\rmA)\big\}$ in Theorem \ref{fixed:known} of the fixed-length test in Algorithm \ref{FA:known}, as justified in Appendix \ref{proof:sequentiality_known}. Thus, there exists the benefit of sequentiality. To illustrate, in Fig. \ref{known_exponent}, we plot the achievable misclassification exponents in Theorems \ref{fixed:known} and \ref{th:known} for the low-complexity fixed-length test in Algorithm \ref{FA:known} and the sequential test in Algorithm \ref{CA:known} when the scoring function $f(\cdot)$ is the GJS divergence, $P_\rmN=\rm{Bern}(0.5)$ and $P_\rmA=\rm{Bern}(a)$ for $a\in(0,1)$ such that $a\neq 0.5$. We choose thresholds for our sequential test as $\lambda_1=0.0005$ and $\lambda_2=f(P_\rmA,P_\rmN)-0.0001$ for each $a$. As shown in Fig. \ref{known_exponent}, our sequential test in Algorithm \ref{CA:known} achieves larger misclassification exponent than fixed-length test in Algorithm \ref{FA:known}.

\begin{figure}[tb]
\centering
\includegraphics[width=0.5\linewidth]{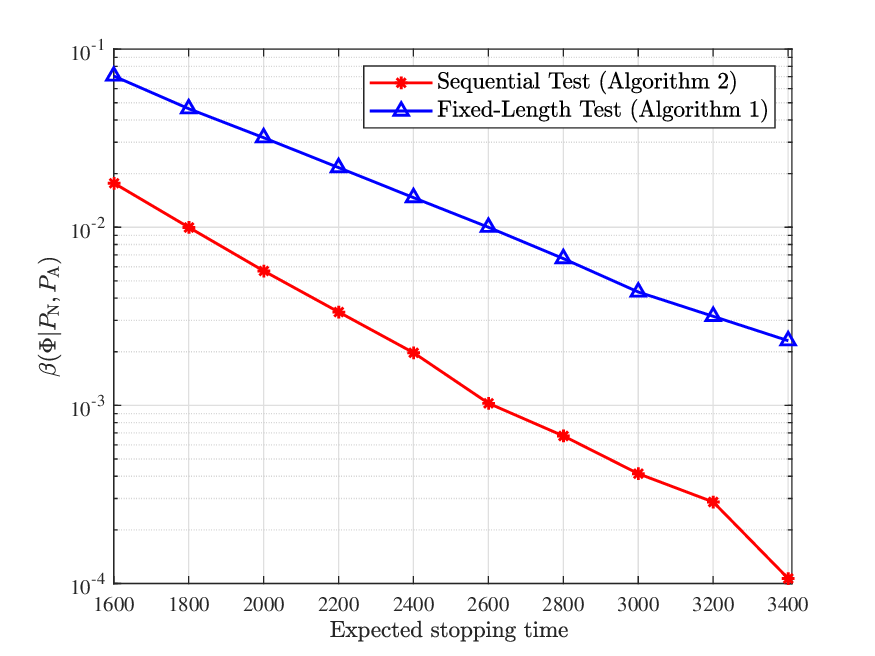}
\caption{Plot of the simulated misclassification probabilities as a function of expected stopping times for the sequential test in Algorithm \ref{CA:known} and fixed-length test in Algorithm \ref{FA:known} when $M=100$, $t=10$, the scoring function $f(\cdot)$ is the GJS divergence, $(P_\rmN,P_\rmA)=\mathrm{Bern}(0.32,0.25)$ and $(\lambda_1,\lambda_2)=(0.001,0.003)$. As observed, our sequential test achieves smaller misclassification probability than the fixed-length test.}
\label{low_compare}
\end{figure}

Furthermore, we numerically illustrate the benefit of sequentiality. Specifically, in Fig. \ref{low_compare}, we plot the simulated misclassification probability for the sequential test in Algorithm \ref{CA:known} and fixed-length test in Algorithm \ref{FA:known} when $M=100$, $t=10$, the scoring function $f(\cdot)$ is the GJS divergence, $(P_\rmN,P_\rmA)=\mathrm{Bern}(0.32,0.25)$ and $(\lambda_1,\lambda_2)=(0.001,0.003)$. As observed, our sequential test performs better than the fixed-length test.

\begin{figure}[tb]
\centering
\includegraphics[width=.5\linewidth]{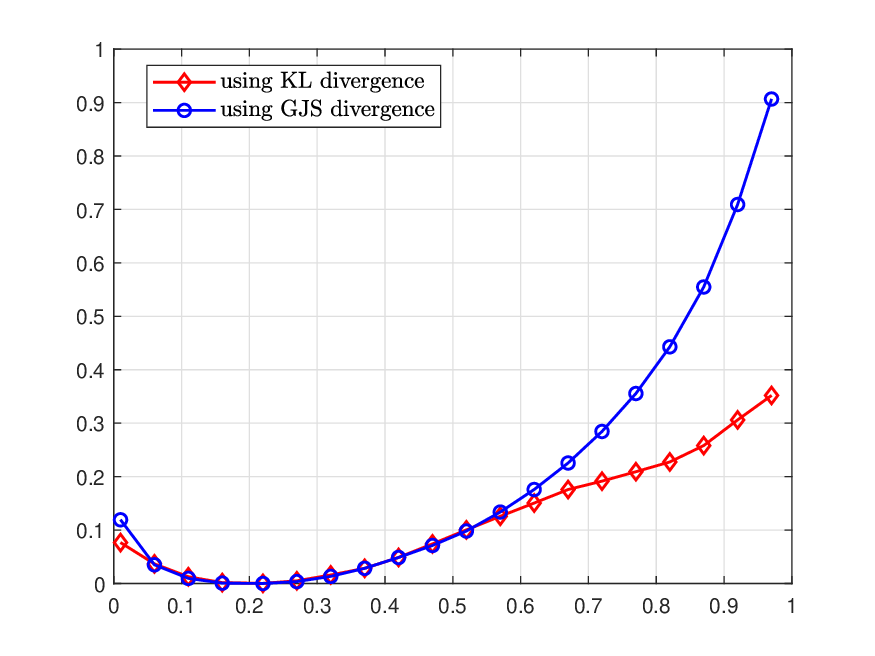}
\caption{Numerical comparison of achievable misclassification exponents of our test in Theorem \ref{th:known} under KL and GJS divergence when $P_\rmN=\rm{Bern}(0.2)$ and $P_\rmA=\rm{Bern}(a)$ for different values of $a\in[0.01,0.99]$ and $a\neq 0.2$, with thresholds $\lambda_1=0.001$ and $\lambda_2=f(P_\rmA,P_\rmN)-0.0001$ for each $a$. As observed, GJS divergence can yield better performance in certain cases.}
\label{kl_vs_gjs_seq}
\end{figure}

Thirdly, we numerically compare the achievable misclassification exponent in Theorem \ref{th:known} of our sequential test for KL and GJS divergences scoring functions. In Fig. \ref{kl_vs_gjs_seq}, we plot the misclassification exponent when $(\lambda_1,\lambda_2)=(0.01,0.02)$, $P_\rmN=\rm{Bern}(0.2)$, and $P_\rmA=\rm{Bern}(a)$ for $a\in[0.01,0.99]$ such that $a\neq 0.2$. As observed, the GJS divergence scoring function generally yields larger misclassification exponent.

\section{Main Results for the Case With Unknown Number of Outliers}

\subsection{Low Complexity Fixed-length Test}

\subsubsection{Test Design and Asymptotic Intuition}
This subsection presents our low-complexity fixed-length test $\Phi_{\mathrm{fix}}^{\rmu}$ when the number of outliers is unknown. Our test generalizes \cite[Algorithm 3]{bu2019linear} by adding an outlier detection phase to deal with the zero outlier case and by allowing the scoring function to be beyond KL divergence.

Define the following set of distinct pair of integers
\begin{align}\label{M_dis}
\calM_{\mathrm{dis}}:=\{(i,j)\in[M]^2:i\neq j\}.
\end{align}
Fix any positive real number $\lambda\in\bbR_+$.  Our test is summarized in Algorithm \ref{FA:unknown} and consists of two phases: outlier detection in steps 1-3 and outlier identification in the remaining steps. In outlier detection, our test calculates all pairwise scoring functions and claims no outlier only if the maximal scoring function value is smaller than the threshold $\lambda$. Otherwise, our test proceeds to outlier detection. In this phase, our test chooses two cluster centers: the first one $c_1$ is chosen at random while the second one $c_2$ is chosen as the type that has largest scoring function value with respect to $c_1$. Subsequently, our test applies binary classification using the minimal scoring function decision rule to form two clusters $\calC_1$ and $\calC_2$. Finally, the indices of outliers are determined as the cluster with smaller size.

\begin{algorithm}[tb]
\caption{Low complexity fixed-length test $\Phi_{\mathrm{fix}}^\rmu$ with unknown number of outliers}
\label{FA:unknown}
\begin{algorithmic}[1]
\Require $M$ observed sequences $\bx^n_M$ and a positive threshold $\lambda\in\bbR_+$
\Ensure A hypothesis in the set $\{\{\rmH_\calB\}_{\calB\in\calS},\mathrm{H_r}\}$
\State Compute
$f\big(\hatT_{x_i^n},\hatT_{x_j^n}\big)$ for all $(i,j)\in\calM_{\rm{dis}}$
\If{$\max_{(i,j)\in\calM_{\rm{dis}}}f\big(\hatT_{x_i^n},\hatT_{x_j^n}\big)\le\lambda$}
\State \Return Hypothesis $\rmH_\rmr$
\Else
\State Choose a number $l\in[M]$ randomly
\State Calculate $i^*=\argmax_{i\in[M]}f\big(\hatT_{x_i^n},\hatT_{x_l^n}\big)$
\State Set $c_1=\hatT_{x_{l}^n}$ and $c_2=\hatT_{x_{i^*}^n}$
\State Set $\calC_1\leftarrow\emptyset$ and $\calC_2\leftarrow\emptyset$
\For{$i\in[M]$}
\State Calculate $k^*=\argmin_{k\in[2]}f\big(\hatT_{x_i^n},c_k\big)$
\State Set $\calC_{k^*}\leftarrow \calC_{k^*}\cup\{i\}$
\EndFor
\State Calculate $t^*=\argmin_{k\in[2]}|\calC_k|$
\State \Return Hypothesis $\rmH_{\calC_{t^*}}$
\EndIf
\end{algorithmic}
\end{algorithm}

We next explain the asymptotic intuition why the above test works. As discussed in Sec. \ref{sec:known_fix}, it follows from the weak law of large numbers that for any $(i,j)\in[M]^2$ such that $i\neq j$, the scoring function $f\big(\hatT_{X_i^n},\hatT_{X_j^n}\big)$ converges to zero if $(X_i^n,X_j^n)$ are both outliers or nominal samples while $f\big(\hatT_{X_i^n},\hatT_{X_j^n}\big)$ converges to a positive real number if there is a nominal sample and an outlier. In outlier detection, if there is no outlier, all the scoring functions $f\big(\hatT_{x_i^n},\hatT_{x_j^n}\big)$ converge to zero and the correct decision of $\rmH_\rmr$ is output for any positive threshold $\lambda$. On the other hand, if there exists an outlier, there exists a scoring function $f\big(\hatT_{x_i^n},\hatT_{x_j^n}\big)$ that is larger than $\lambda$ when $\lambda<\min\{f(P_\rmA,P_\rmN),~f(P_\rmN,P_\rmA)\}$ and the test proceeds to outlier detection. In outlier detection, following the same logic, with asymptotically probability one, the cluster centers $c_1$ and $c_2$ correspond to types of a nominal sample and an outlier although it is not certain whether $c_1$ or $c_2$ corresponds to an outlier. Similarly, the clusters $\calC_1$ and $\calC_2$ collect indices of nominal samples and outliers, respectively. Finally, the correct index set of outliers can be identified as the set $\calC_{t^*}$ that has smaller size between $(\calC_1,\calC_2)$ because the number of outliers is smaller than the number of nominal samples.

\subsubsection{Theoretical Results and Discussions}
Fix any pair of distributions $(P_1,P_2)\in\calP(\calX)^2$. Define the following exponent function
\begin{align}\label{gamma}
\gamma(P_1,P_2):=\min_{\substack{(Q_1,Q_2,Q_3)\in\calP(\calX)^3:f(Q_1,Q_3)\le f(Q_1,Q_2)}}D(Q_1||P_1)+D(Q_2||P_1)+D(Q_3||P_2).
\end{align}
Recall the definitions of exponent functions of $\eta(P_1,P_2)$ in \eqref{eta0}, $\Omega(P_1,P_2,\lambda)$ in \eqref{omega} and $\Upsilon(P,\lambda)$ in \eqref{upsilon}.


\begin{theorem}\label{fixed:unknown}
Given any $\lambda\in\bbR_+$, under any pair of distributions $(P_\rmN,P_\rmA)\in\calP(\calX)^2$, the fixed-length test in Algorithm \ref{FA:unknown} ensures that
\begin{itemize}
\item for each $\calB\in\calS$,
\begin{itemize}
\item the misclassification exponent satisfies
\begin{align}
E_{\beta_\calB}(\Phi_{\mathrm{fix}}^{\rmu}|P_\rmN,P_\rmA)\ge\min\big\{\eta(P_\rmN,P_\rmA),~\eta(P_\rmA,P_\rmN),~\gamma(P_\rmA,P_\rmN),~\gamma(P_\rmN,P_\rmA)\big\}.
\end{align}
\item the false reject exponent satisfies
\begin{align}
E_{\zeta_\calB}(\Phi_{\mathrm{fix}}^{\rmu}|P_\rmN,P_\rmA)\ge\max\{\Omega(P_\rmA,P_\rmN,\lambda),~\Omega(P_\rmN,P_\rmA,\lambda)\}.
\end{align}
\end{itemize}
\item the false alarm exponent satisfies
\begin{align}
E_{\rm fa}(\Phi_{\mathrm{fix}}^{\rmu}|P_\rmN,P_\rmA)\ge\Upsilon(P_\rmN,\lambda).
\end{align}
\end{itemize}
\end{theorem}
The proof of Theorem \ref{fixed:unknown} is provided in Appendix \ref{proof:fix_unknown}.
The misclassification exponent is lower bounded by the minimization of four exponent functions. The results are obtained by analyzing the exponential decay rates of two error events: i) $\calE^\mathrm{f,u}_1$ where in step 7, $c_1$ and $c_2$ are types of either two outliers or two nominal samples, and ii) $\calE^\mathrm{f,u}_2$ where in steps 10-11, an outlier is incorrectly identified as a nominal sample or a nominal sample is incorrectly classified as an outlier when $(\calE^\mathrm{f,u}_1)^\rmc$ occurs. In particular, $\eta(P_\rmN,P_\rmA)$ characterizes the exponential decay rates for the probability of the error event $\calE^\mathrm{f,u}_1$ when both cluster centers are types of nominal samples while $\eta(P_\rmA,P_\rmN)$ characterizes the exponential decay rates for the probability of the error event $\calE^\mathrm{f,u}_1$ when both cluster centers are types of outliers. Analogously,
$\gamma(P_\rmA,P_\rmN)$ characterizes the exponential decay rate for the probability of $\calE^\mathrm{f,u}_2$ where an outlier is classified as a nominal sample while $\gamma(P_\rmN,P_\rmA)$ characterizes the exponential decay rate for the probability of $\calE^\mathrm{f,u}_2$ where a nominal sample is classified as an outlier.

We make several remarks. Firstly, the threshold $\lambda$ trades off the false reject and false alarm exponents. Specifically, the false reject exponent $\max\{\Omega(P_\rmA,P_\rmN,\lambda),~\Omega(P_\rmN,P_\rmA,\lambda)\}$ is non-increasing in $\lambda$ while the false alarm exponent $\Upsilon(P_\rmN,\lambda)$ is non-decreasing in $\lambda$. Note that the false reject exponent lower bounds the exponential decay rate for the probability that the maximal pairwise scoring function is below the threshold $\lambda$ when there exists at least one outlier while the false alarm exponent lower bounds the exponential decay rate for the probability that the maximal pairwise scoring function is above the threshold $\lambda$ when there is no outlier. Furthermore, the false alarm exponent $\Upsilon(P_\rmN,\lambda)$ is always positive for any $\lambda\in\bbR_+$ while the false reject exponent  $\max\{\Omega(P_\rmA,P_\rmN,\lambda),~\Omega(P_\rmN,P_\rmA,\lambda)\}$ is strictly positive if $\lambda<\max\{f(P_\rmA,P_\rmN),~f(P_\rmN,P_\rmA)\}$.

Secondly, comparing Theorems \ref{fixed:known} and \ref{fixed:unknown}, we reveal the penalty of not knowing the number of outliers on the performance of low-complexity fixed-length tests under non-null hypotheses. Recall that in Theorem \ref{fixed:known}, it is assumed that $t$ outliers exist while in Theorem \ref{fixed:unknown}, the number of outliers is unknown but upper bounded by an integer $T$. For fair comparison, we should consider the error probability under each non-null hypothesis. This corresponds to compare the misclassification exponent in Theorem \ref{fixed:known}, i.e, $E_\calB(\Phi_{\mathrm{fix}}|P_\rmN,P_\rmA)$, with the minimal value of the misclassification and the false reject exponents in Theorem \ref{fixed:unknown}, i.e., $\min\{E_{\beta_\calB}(\Phi_{\mathrm{fix}}|P_\rmA,P_\rmN),~E_{\zeta_\calB}(\Phi_{\mathrm{fix}}|P_\rmA,P_\rmN)\}$.
It follows that
\begin{align}
\nn&\min\{\eta(P_\rmA,P_\rmN),~\eta(P_\rmN,P_\rmA)\}\\
&\ge\min\big\{\eta(P_\rmA,P_\rmN),~\eta(P_\rmN,P_\rmA),~\gamma(P_\rmA,P_\rmN),~\gamma(P_\rmN,P_\rmA),
~\Omega(P_\rmA,P_\rmN,\lambda),~\Omega(P_\rmN,P_\rmA,\lambda)\big\}\label{fl:penalty}.
\end{align}
Thus, the fixed-length test that knows the number of outliers has better performance than the fixed-length test that does not know the number of outliers. In the following numerical example, we show that the penalty can be strict. When the scoring function $f(\cdot)$ is the GJS divergence, $(P_\rmN,P_\rmA)=\mathrm{Bern}(0.4,0.9)$ and $\lambda=0.08$, it follows that $\min\{\eta(P_\rmA,P_\rmN),~\eta(P_\rmN,P_\rmA)\}=0.107$, which is strictly greater than $0.0823$ of the right hand side of \eqref{fl:penalty}.

\begin{figure}[tb]
\centering
\includegraphics[width=0.5\linewidth]{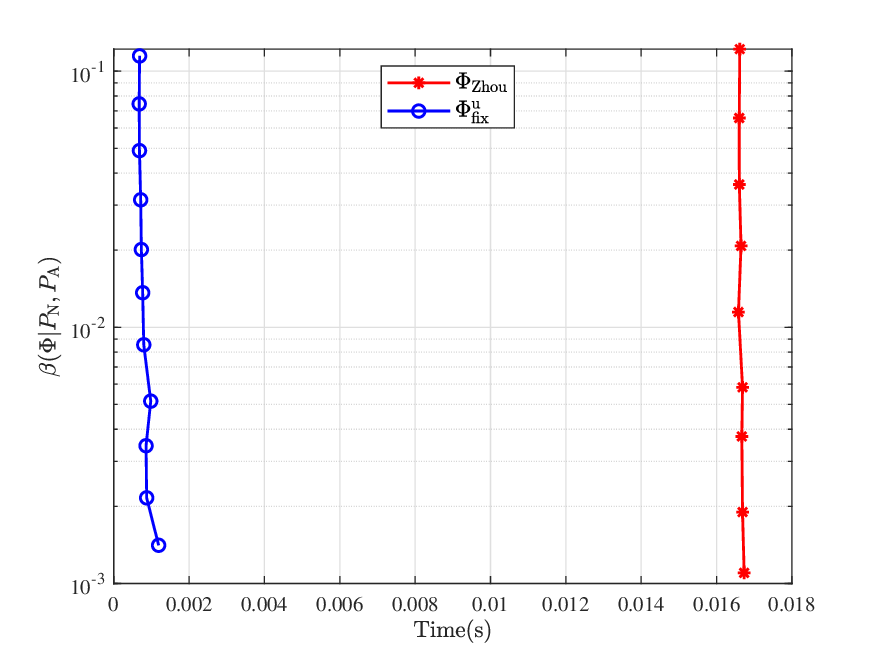}
\caption{
Plot of simulated misclassification probabilities as a function of running times for our test in Algorithm \ref{FA:unknown} and the fixed-length test $\Phi_{\rm Zhou}$ in \eqref{test_zhou} under distributions $(P_\rmN,P_\rmA)=\mathrm{Bern}(0.23,0.3)$ with threshold $\lambda=0.001$ when $M=10$, $T=4$, $|\calB|=3$ and $f(\cdot)$ is the GJS divergence. As observed, our test achieves the same misclassification probability with much less running time than the test $\Phi_{\rm Zhou}$.
}
\label{comp:zhou}
\end{figure}

Finally, to reveal the advantage of computational complexity of the test in Algorithm \ref{FA:unknown}, in Fig. \ref{comp:zhou}, we numerically compare our low-complexity fixed-length test in Algorithm \ref{FA:unknown} and the exhaustive search fixed-length test $\Phi_{\rm Zhou}$ in \eqref{test_zhou} when $M=10$, $T=4$, $|\calB|=3$ $(P_\rmN,P_\rmA)=\mathrm{Bern}(0.23,0.3)$ and there are three outliers. For our test, the scoring function $f(\cdot)$ is the GJS divergence and the threshold is $\lambda=0.001$. As observed in Fig. \ref{comp:zhou}, our test achieves a much better tradeoff between detection performance and computational complexity.

\subsection{Low Complexity Sequential Test}
\subsubsection{Test Design and Asymptotic Intuition}
This subsection presents our low-complexity sequential test when the number of outliers is unknown. Recall the definition of $\calM_{\mathrm{dis}}$ in \eqref{M_dis}. Given parameters $(\lambda_1,\lambda_2,n)\in\bbR_+^2\times\bbN$ such that $\lambda_1\leq \lambda_2$, our sequential test $\Phi_\mathrm{seq}^\rmu=(\tau,\phi)$ is summarized in Algorithm \ref{CA:unknown}.

\begin{algorithm}[tb]
\caption{Low complexity sequential test $\Phi_{\mathrm{seq}}^\rmu$ with unknown number of outliers}
\label{CA:unknown}
\begin{algorithmic}[1]
\Require $M$ observed sequences and two thresholds $(\lambda_1,\lambda_2)\in\bbR_+^2$ such that $\lambda_1\leq \lambda_2$
\Ensure A stopping time $\tau$ and a hypothesis $\hat{\rmH}$ in the set $\{\{\rmH_\calB\}_{\calB\in\calS},\mathrm{H_r}\}$
\State Set $k=n-1$ and initialize $\mathrm{flag}=0$
\State Collect samples $(x_1^k,\ldots,x_M^k)$.
\While{$\mathrm{flag}=0$}
\State Compute
$f\big(\hatT_{x_i^k},\hatT_{x_j^k}\big)$ for all $(i,j)\in\calM_{\rm{dis}}$
\If{$\max_{(i,j)\in\calM_{\rm{dis}}}f\big(\hatT_{x_i^k},\hatT_{x_j^k}\big)\le\lambda_1$}
\State Set $\mathrm{flag}=1$
\State \Return $\tau=k$ and $\hat{\rmH}=\rmH_\rmr$
\State break
\EndIf
\If{$\max_{(i,j)\in\calM_{\rm{dis}}}f\big(\hatT_{x_i^k},\hatT_{x_j^k}\big)>\lambda_2$}
\State Choose a number $l\in[M]$ randomly and set $\hatT_0=\hatT_{x_l^k}$
\State Set $\calC_1=\emptyset$ and $\calC_2=\emptyset$
\For {$i\in[M]$}
\State Compute $f\big(\hatT_{x_i^k},\hatT_0\big)$
\If{$f\big(\hatT_{x_i^k},\hatT_0\big)<\lambda_1$}
\State $\calC_1\leftarrow \calC_1\cup\{i\}$
\ElsIf {$f\big(\hatT_{x_i^k},\hatT_0\big)>\lambda_2$}
\State $\calC_2\leftarrow \calC_2\cup\{i\}$
\EndIf
\EndFor
\If {$|\calC_1|+|\calC_2|=M$}
\State Calculate $t^*=\argmin_{k\in[2]}|\calC_k|$
\State Set $\mathrm{flag}=1$
\State \Return $\tau=k$ and output $\hat{\rmH}=\rmH_{\calC_{t^*}}$
\State break
\EndIf
\EndIf
\State Collect new symbols $(x_{1,k+1},\ldots,x_{M,k+1})$
\State Update $k$ as $k+1$
\EndWhile
\end{algorithmic}
\end{algorithm}

Similar to the sequential test in Algorithm \ref{CA:known}, our sequential test has the minimal stopping time and initializes the sample size as $k=n-1$. Similar to the fixed-length test in Algorithm \ref{FA:unknown}, our low-complexity sequential test in Algorithm \ref{CA:unknown} consists of two phases: outlier detection in steps 5-8 and outlier identification in the remaining steps. In outlier detection, our test calculates all pairwise scoring functions, claims no outlier if the maximal value is smaller than $\lambda_1$, and proceeds to outlier detection phase if the maximal value is larger than $\lambda_2$. If the maximal value is between $\lambda_1$ and $\lambda_2$, our test collects new samples and iterates. Once the test proceeds to the outlier identification phase, the test randomly chooses a sequence and sets its type as $\hatT_0$. Subsequently, our test classifies each sequence as either a nominal sample or an outlier with two sets $(\calC_1,\calC_2)$ using binary classification with thresholds $(\lambda,\lambda_2)$. If all sequences are classified reliably, the test stops and claims the indices of outliers as the set with smaller size between two sets $(\calC_1,\calC_2)$.

Our sequential low-complexity test has much lower computational complexity than the existing sequential test $\Phi_{\rm Diao}$ in \eqref{test:unknown}. Specifically, our test utilizes the pairwise scoring function to find the outlier set, which incurs polynomial complexity with respect to the number of sequences $M$, regardless of the number of outliers. In contrast, the existing test $\Phi_{\rm Diao}$ in \eqref{test:unknown} applies exhaustive search, whose computational complexity is proportional to $\sum_{i=1}^{T}{M\choose i}$ and could be prohibitively large for relatively large numbers $M$ and $T$.

We next explain the asymptotic intuition why the above test works. The outlier detection phase follows the same asymptotic intuition as Algorithm \ref{FA:unknown}. In particular, if there is no outlier, all the scoring functions converge to zero and the correct decision of $\rmH_\rmr$ is output for any positive $\lambda_1$. On the other hand, if there exists an outlier, there exists a scoring function that is larger than $\lambda_2$ for any $0<\lambda_2<\min\{f(P_\rmA,P_\rmN),~f(P_\rmN,P_\rmA)\}$, so that the test proceeds to the outlier identification phase. The outlier identification phase is essentially binary classification as in Algorithm \ref{CA:known}, which shares the same asymptotic intuition and thus omitted.

In the next subsection, we characterize the achievable large deviations performance of the sequential test in Algorithm \ref{CA:unknown}.

\subsubsection{Theoretical Results and Discussions}
Recall the definitions of error exponent functions of $\Omega(P_1,P_2,\lambda)$ in \eqref{omega} and $\Upsilon(P,\lambda)$ in \eqref{upsilon}.
\begin{theorem}\label{th:unknown}
Under any pair of distributions $(P_\rmN,P_\rmA)\in\calP(\calX)^2$, given any parameters $(\lambda_1,\lambda_2)\in\bbR_+^2$ such that $0<\lambda_1\leq \lambda_2<\min\{f(P_\rmA,P_\rmN),~f(P_\rmN,P_\rmA)\}$, our sequential test in Algorithm \ref{CA:unknown} satisfies the expected stopping time universality constraint and ensures that
\begin{itemize}
\item for each $\calB\in\calS$,
\begin{itemize}
\item the misclassification exponent satisfies
\begin{align}
E_{\beta_\calB}(\Phi_\mathrm{seq}^{\rmu}|P_\rmN,P_\rmA)\ge\min\big\{\Omega(P_\rmN,P_\rmA,\lambda_1),~
\Omega(P_\rmA,P_\rmN,\lambda_1),~\Upsilon(P_\rmN,\lambda_2),~\Upsilon(P_\rmA,\lambda_2)\big\}.
\end{align}
\item the false reject exponent satisfies
\begin{align}
E_{\zeta_\calB}(\Phi_\mathrm{seq}^{\rmu}|P_\rmN,P_\rmA)&\ge\max\{\Omega(P_\rmA,P_\rmN,\lambda_1),~\Omega(P_\rmN,P_\rmA,\lambda_1)\}.
\end{align}
\end{itemize}
\item the false alarm exponent satisfies
\begin{align}
E_{\rm fa}(\Phi_\mathrm{seq}^{\rmu}|P_\rmN,P_\rmA)&\ge\Upsilon(P_\rmN,\lambda_2).
\end{align}
\end{itemize}
\end{theorem}
The proof of Theorem \ref{th:unknown} is provided in \ref{proof:unknown}.
The misclassification exponent is lower bounded by the minimization of four exponent functions. The results are obtained by analyzing the exponential decay rates of following misclassification error events: i) $\calE^\mathrm{s,u}_1$ where a nominal sample is falsely identified as an outlier and ii) $\calE^\mathrm{s,u}_2$ where an outlier is falsely identified as a nominal sample. In particular, $\Upsilon(P_\rmN,\lambda_2)$ and $\Omega(P_\rmN,P_\rmA,\lambda_1)$ lower bound the exponential decay rates for the probabilities of the error event $\calE^\mathrm{s,u}_1$ when $\hatT_0$ chosen randomly in step 9 is the type of a nominal sample and an outlier, respectively. Analogously, $\Omega(P_\rmA,P_\rmN,\lambda_1)$ and $\Upsilon(P_\rmA,\lambda_2)$ lower bound the exponential decay rates for the probabilities of the error event $\calE^\mathrm{s,u}_2$ when $\hatT_0$ is the type of a nominal sample and an outlier, respectively.

We make several remarks. Firstly, comparing Theorems \ref{th:known} and \ref{th:unknown}, we reveal the penalty of not knowing the number of outliers on the performance of sequential tests. Recall that in Theorem \ref{th:known}, it is known that $t$ outliers exist while in Theorem \ref{th:unknown}, the number of outliers is unknown, which can be any number from $0$ to $T$. For fair comparison, we should consider the error probabilities under each non-null hypothesis and thus compare the misclassification exponent $\min\big\{\Omega(P_\rmN,P_\rmA,\lambda_1),~\Upsilon(P_\rmN,\lambda_2)\big\}$ in Theorem \ref{th:known} with the Bayesian exponent in Theorem \ref{th:unknown}, which is given by the minimal value of the misclassification and the false reject exponents, i.e., $\min\{\Omega(P_\rmN,P_\rmA,\lambda_1),~
\Omega(P_\rmA,P_\rmN,\lambda_1),~\Upsilon(P_\rmN,\lambda_2),~\Upsilon(P_\rmA,\lambda_2)\}$. For any $(\lambda_1,\lambda_2)\in\bbR_+^2$ such that $\lambda_1\le\lambda_2$, it follows that
\begin{align}
\min\big\{\Omega(P_\rmN,P_\rmA,\lambda_1),~\Upsilon(P_\rmN,\lambda_2)\}
&\ge\min\{\Omega(P_\rmN,P_\rmA,\lambda_1),~
\Omega(P_\rmA,P_\rmN,\lambda_1),~\Upsilon(P_\rmN,\lambda_2),~\Upsilon(P_\rmA,\lambda_2)\}.
\end{align}
Thus, there is a penalty on the achievable exponent when the number of outlier is unknown. In the following, we numerically show that such penalty can be strict. Set the scoring function $f(\cdot)$ as the GJS divergence. When $(P_\rmN,P_\rmA)=\mathrm{Bern}(0.4,0.9)$ and $(\lambda_1,\lambda_2)=(0.06,0.08)$, it follows that $\min\{\Omega(P_\rmN,P_\rmA,\lambda_1),~\Upsilon(P_\rmN,\lambda_2)\}=0.0827$ while $\min\{\Omega(P_\rmN,P_\rmA,\lambda_1),~
\Omega(P_\rmA,P_\rmN,\lambda_1),$ $\Upsilon(P_\rmN,\lambda_2),~\Upsilon(P_\rmA,\lambda_2)\}=0.0807$.

\begin{figure}[tb]
\centering
\subfloat[Misclassification Exponent]{
\label{mis_compare}
\includegraphics[width=0.45\linewidth]{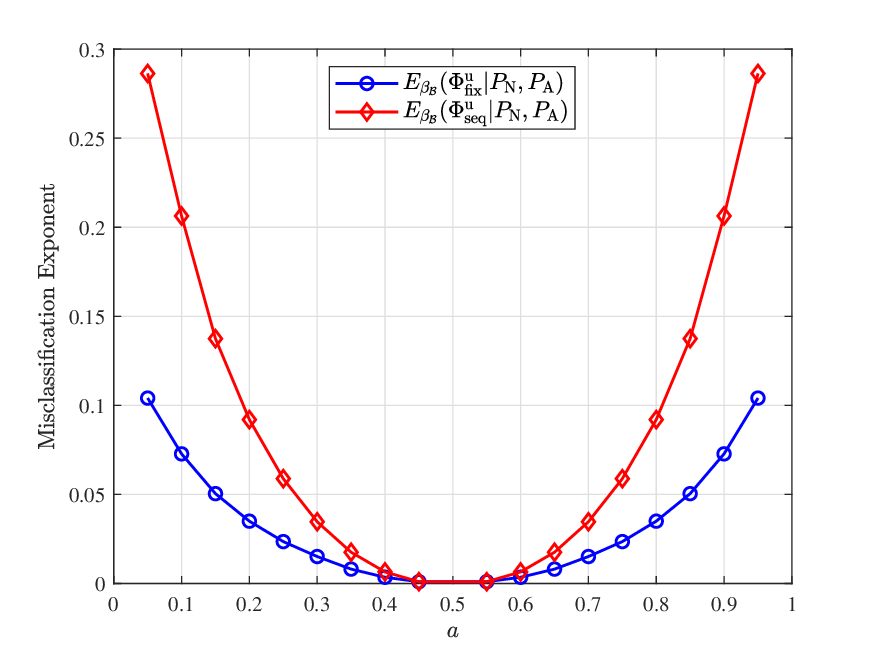}
}\hfill
\subfloat[False Reject Exponent]{
\label{fr_compare}
\includegraphics[width=0.45\linewidth]{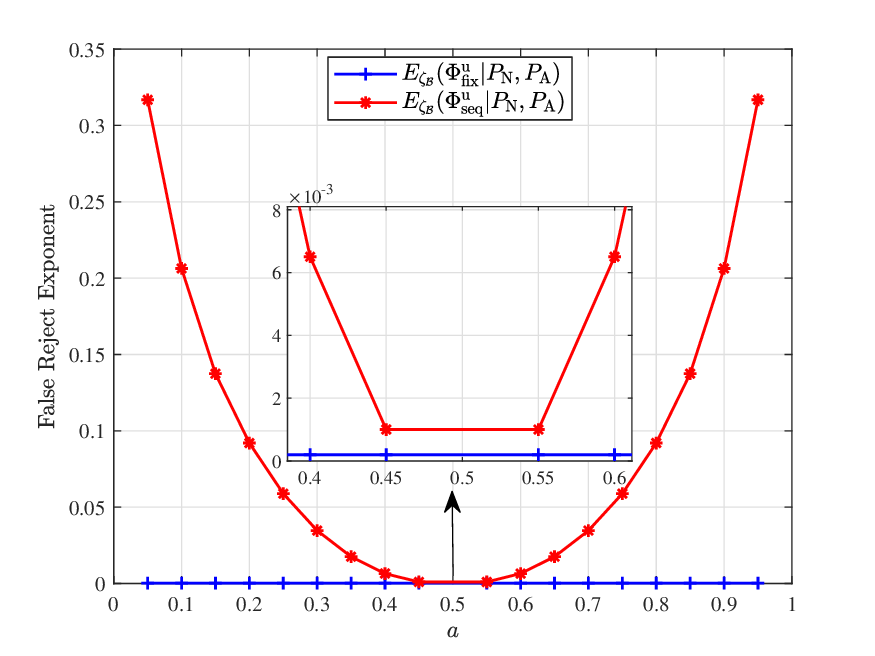}
}
\caption{Plot of achievable misclassification and false reject exponents for the sequential test in Theorem \ref{th:unknown} and the fixed-length test in Theorem \ref{fixed:unknown} when the scoring function $f(\cdot)$ is the GJS divergence, $P_\rmN=\rm{Bern}(0.5)$, $P_\rmA=\rm{Bern}(a)$ for $a\in(0,1)$ such that $a\neq 0.5$, with thresholds $\lambda_1=0.0005$ and $\lambda=\lambda_2=f(P_\rmA,P_\rmN)-0.0001$ for each $a$. As observed, both exponents for the sequential test are larger than that for the fixed-length test.}
\label{unknown_exponent}
\end{figure}

Secondly, comparing Theorems \ref{fixed:unknown} and \ref{th:unknown}, we reveal the benefit of sequentiality in terms of the Bayesian error exponent, which is the minimal value of achievable misclassification and false reject exponents when the false alarm exponents of both cases are the same.  The justification is provided in Appendix \ref{proof:sequentiality_unknown}. To illustrate, in Fig. \ref{unknown_exponent}, we plot the achievable misclassification and false reject exponents in Theorems \ref{fixed:unknown} and \ref{th:unknown} for the low-complexity fixed-length test in Algorithm \ref{FA:unknown} and the sequential test in Algorithm \ref{CA:unknown} when the scoring function $f(\cdot)$ is the GJS divergence, $P_\rmN=\rm{Bern}(0.5)$ and $P_\rmA=\rm{Bern}(a)$ for $a\in(0,1)$ such that $a\neq 0.5$. We choose thresholds for our sequential test as $\lambda_1=0.0005$ and $\lambda=\lambda_2=f(P_\rmA,P_\rmN)-0.0001$ for each $a$. Since $\lambda=\lambda_2$, the false alarm exponents for both tests are the same. As shown in Fig. \ref{unknown_exponent}, our sequential test in Algorithm \ref{CA:known} achieves larger misclassification and false reject exponents than fixed-length test in Algorithm \ref{FA:unknown}. 

\begin{figure}[tb]
\centering
\includegraphics[width=0.5\linewidth]{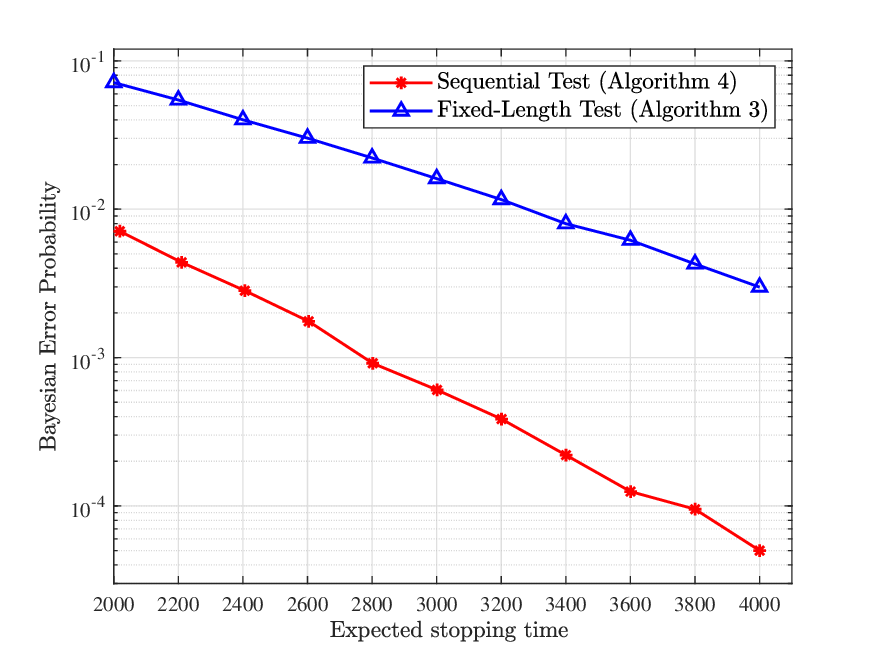}
\caption{Plot of the simulated Bayesian probability as a function of expected stopping times for the sequential test in Algorithm \ref{CA:unknown} and fixed-length test in Algorithm \ref{FA:unknown} when $M=100$, $T=20,~|\calB|=10$, the scoring function $f(\cdot)$ is the GJS divergence, $(P_\rmN,P_\rmA)=\mathrm{Bern}(0.32,0.25)$ and $\lambda_1=0.001,\lambda=\lambda_2=0.0025$. As observed, our sequential tests achieves smaller Bayesian probability than the fixed-length test.}
\label{low_compare_unknown}
\end{figure}

Finally, we numerically illustrate the benefit of sequentiality in Fig. \ref{low_compare_unknown}. Specifically,  we plot the simulated Bayesian error probabilities under the non-null hypothesis, which is the weighted sum of misclassification and false reject probabilities, for the sequential test in Algorithm \ref{CA:unknown} and fixed-length test in Algorithm \ref{FA:unknown} when $M=100$, $T=20,~|\calB|=10$, the scoring function $f(\cdot)$ is GJS divergence, $(P_\rmN,P_\rmA)=\mathrm{Bern}(0.32,0.25)$ and $(\lambda_1,\lambda_2,\lambda)=(0.001,0.0025,0.0025)$. As observed, our sequential test is superior to the fixed-length test by achieving smaller Bayesian error probabilities.

\section{Conclusion}

We revisited outlier hypothesis testing and proposed low-complexity exponentially consistent fixed-length and sequential tests when the nominal and anomalous distributions are unknown and when the number of outliers is either known and unknown. In particular, our sequential tests have bounded expected stopping times and all our low-complexity tests incur polynomial complexity with respect to the total number of observed sequences regardless of the number of outliers. Compared with the optimal tests in \cite{li2014,zhou2022second,diao2025sequential} that use exhaustive search and incur forbiddingly high computational complexity, our low-complexity tests strike a better tradeoff between detection performance and computational complexity. Furthermore, comparing our results for the case with known and unknown number of outliers, we reveal the penalty of not knowing the number of outliers on the performance of both fixed-length and sequential tests. Comparing our results for fixed-length and sequential tests, we reveal the benefit of sequentiality. Our results are illustrated via numerical examples.

We next discuss future directions. Firstly, we assumed all nominal samples are generated from the same nominal distribution and all outliers are generated from the same anomalous distribution. However, in practice, nominal samples could be generated from different distributions that deviate slightly, so are the outliers. Thus, towards a further step of practical applications, it is worthwhile to generalize our results to account for distribution uncertainty, using exponential families~\cite{pan2022asymptotics} or the distribution ball~\cite{huber1965robust,hsu2020binary}. Secondly, we assumed that all observed sequences are discrete. However, in practical applications, the observed sequences can take real values. Thus, it is beneficial to generalize our results to account for continuous observed sequences, potentially using the kernel methods~\cite{gretton2012kernel,zou2017nonparametric,zhu2024exponentially}. Finally, it would be of great interest to generalize the ideas of constructing low-complexity tests in this paper to other statistical inference problems, e.g., clustering~\cite{efimov2019adaptive}, statistical sequence matching~\cite{unnikrishnan2014asymptotically,zhou2024large}, and quickest change-point detection~\cite{smith1975bayesian,pettitt1979non}.

\appendix

\subsection{Proof of Theorem \ref{fixed:known} (Fixed-length Test with Known Number of Outliers)}
\label{proof:fix_known}
Recall that the number of outliers is $t$ and the fixed-length test $\Phi_{\mathrm{fix}}$ is summarized in Algorithm \ref{FA:known}. Fix any $\calB\in\calS(t)$.

A misclassification event of the test $\Phi_{\mathrm{fix}}$ occurs if one of the following two events occurs: i) $\calE^\mathrm{f,k}_1$ where $\tilP_\rmN$ chosen in step 4 is the type of an outlier, and ii) $\calE^\mathrm{f,k}_2$ where in step 7, the test incorrectly claims a nominal sample as an outlier when $(\calE^\mathrm{f,k}_1)^\rmc$ occurs. The error event $\calE^\mathrm{f,k}_1$ can be further categorized into two events: $\calE^\mathrm{f,k}_{1,1}$ where $\calE^\mathrm{f,k}_1$ occurs when $\hatT_0$ chosen in step 1 is the type of a nominal sample and $\calE^\mathrm{f,k}_{1,2}$ where $\calE^\mathrm{f,k}_1$ occurs when $\hatT_0$ is the type of an outlier.

It follows from the test design in Algorithm \ref{FA:known} that the event $(\calE^\mathrm{f,k}_{1,1})^\rmc$ occurs if the scoring function between the type of any outlier and $\hatT_0$ is greater than the scoring function between the type of any nominal sample and $\hatT_0$.
Therefore, the event $\calE^\mathrm{f,k}_{1,1}$ implies there exists an outlier and a nominal sample such that the scoring function between the type of the outlier and $\hatT_0$ is smaller than the scoring function between the type of the nominal sample and $\hatT_0$. Using the fact that $\hatT_0$ is the type of a nominal sample, the probability of $\calE^\mathrm{f,k}_{1,1}$ can be upper bounded by the probability of the following event $\bar\calE^\mathrm{f,k}_{1,1}$:
\begin{align}
\bar\calE^\mathrm{f,k}_{1,1}:=\Big\{\exists~i\in\calB,~\exists~(j_1,j_2)\in(\calM_\calB)^2,~j_1\neq j_2:
f\Big(\hatT_{X_i^n},\hatT_{X_{j_1}^n}\Big)\leq f\Big(\hatT_{X_{j_2}^n},\hatT_{X_{j_1}^n}\Big)\Big\}\label{def:calR1'}.
\end{align}
Analogously, the probability of $\calE^\mathrm{f,k}_{1,2}$ can be upper bounded by the probability of the following event $\bar\calE^\mathrm{f,k}_{1,2}$:
\begin{align}
\bar\calE^\mathrm{f,k}_{1,2}:=\Big\{\exists~(i_1,i_2)\in\calB^2,~i_1\neq i_2,~\exists~j\in\calM_\calB:f\Big(\hatT_{X_j^n},\hatT_{X_{i_1}^n}\Big)\leq f\Big(\hatT_{X_{i_2}^n},\hatT_{X_{i_1}^n}\Big)\Big\}\label{def:calR2'}.
\end{align}
Since $\calE^\mathrm{f,k}_1=\calE^\mathrm{f,k}_{1,1}\cup \calE^\mathrm{f,k}_{1,2}$, it follows that $\Pr\{\calE^\mathrm{f,k}_1\}\le\Pr\{\bar\calE^\mathrm{f,k}_{1,1}\}+\Pr\{\bar\calE^\mathrm{f,k}_{1,2}\}$.

Conditioned on $(\calE^\mathrm{f,k}_1)^\rmc$, the event $\calE^\mathrm{f,k}_2$ occurs if there exists an outlier whose type is closer to $\tilP_\rmN$. It follows that
\begin{align}
\calE^\mathrm{f,k}_2
&=(\calE^\mathrm{f,k}_1)^\rmc\bigcap\Big\{\exists~i\in\calB,~\exists~j\in\calM_\calB:f\Big(\hatT_{x_i^n},\tilP_\rmN\Big)\leq f\Big(\hatT_{X_j^n},\tilP_\rmN\Big)\Big\}\\
&\subseteq(\calE^\mathrm{f,k}_1)^\rmc\bigcap\Big\{\exists~i\in\calB,~\exists~(j_1,j_2)\in(\calM_\calB)^2,~j_1\neq j_2:f\Big(\hatT_{X_i^n},\hatT_{X_{j_1}^n}\Big)\leq f\Big(\hatT_{X_{j_2}^n},\hatT_{X_{j_1}^n}\Big)\Big\}\label{calw:sup}\\
&=(\calE^\mathrm{f,k}_1)^\rmc\cap \bar\calE^\mathrm{f,k}_{1,1}\label{usedef:r1'}\\
&\subset \bar\calE^\mathrm{f,k}_{1,1}.\label{eventW}
\end{align}
where \eqref{calw:sup} follows since $\tilP_\rmN$ is type of a nominal sample when the event $(\calE^\mathrm{f,k}_1)^\rmc$ occurs and \eqref{usedef:r1'} follows from the definition of $\bar\calE^\mathrm{f,k}_{1,1}$ in \eqref{def:calR1'}. Therefore, combining the above analyses, we conclude that the misclassification probability satisfies
\begin{align}
\beta_\calB(\Phi|P_\rmN,P_\rmA)&=\bbP_\calB\{\Phi(\bX^n)\neq\rmH_\calB\}\\
&\leq \bbP_\calB\{\calE^\mathrm{f,k}_1\cup \calE^\mathrm{f,k}_2\}\\
&\leq \bbP_\calB\{\calE^\mathrm{f,k}_1\}+\bbP_\calB\{\calE^\mathrm{f,k}_2\}\\
&\leq \bbP_\calB\big\{\bar\calE^\mathrm{f,k}_{1,1}\big\}+ \bbP_\calB\big\{\bar\calE^\mathrm{f,k}_{1,2}\big\}+\bbP_\calB\big\{\bar\calE^\mathrm{f,k}_{1,1}\big\}\\
&=2\bbP_\calB\big\{\bar\calE^\mathrm{f,k}_{1,1}\big\}+\bbP_\calB\big\{\bar\calE^\mathrm{f,k}_{1,2}\big\}.\label{GF1}
\end{align}

We next bound the probabilities of events $\big(\bar\calE^\mathrm{f,k}_{1,1},\bar\calE^\mathrm{f,k}_{1,2}\big)$. For ease of notation,  define the set
\begin{align}
\calA:=\{(Q_1,Q_2,Q_3)\in\calP(\calX)^3:f(Q_1,Q_2)\le f(Q_3,Q_2)\}\label{def:calA}.
\end{align}
and given any observed sequences $\bx^n=(x_1^n,\ldots,x_M^n)$ and $(i,j,l)\in[M]^3$, let $\bx_{i,j,l}^n:=(x_i^n,x_j^n,x_l^n)$ and let $\hatT_{\bx_{i,j,l}^n}:=(\hatT_{x_i^n},\hatT_{x_j^n}.\hatT_{x_l^n})$.
It follows from the method of types~\cite{csiszar1998mt} that
\begin{align}
\bbP_\calB\big\{\bar\calE^\mathrm{f,k}_{1,1}\big\}
&\le\sum_{i\in\calB}\sum_{\substack{(j_1,j_2)\in(\calM_\calB)^2:\\j_1\neq j_2}}\bbP_\calB\Big\{f\Big(\hatT_{X_i^n},\hatT_{X_{j_1}^n}\Big)
\le f\Big(\hatT_{X_{j_2}^n},\hatT_{X_{j_1}^n}\Big)\Big\}\label{sym1}\\
&\le \sum_{i\in\calB}\sum_{\substack{(j_1,j_2)\in(\calM_\calB)^2:\\j_1\neq j_2}}\sum_{\substack{\bx_{i,j_1,j_2}^n\in\calX^{3n}:\\\hatT_{\bx_{i,j_1,j_2}^n}\in\calA}}
P_\rmA\big(x^n_i\big)P_\rmN\big(x_{j_1}^n\big) P_\rmN\big(x_{j_2}^n\big)\label{numberB}\\
&\le \sum_{i\in\calB}\sum_{\substack{(j_1,j_2)\in(\calM_\calB)^2:\\j_1\neq j_2}}\sum\limits_{\substack{\bQ\in\calA}}
P_\rmA(\calT_{Q_1}^n)P_\rmN(\calT_{Q_2}^n) P_\rmN(\calT_{Q_3}^n)\\
&\le t(M-t)^2\sum\limits_{\substack{\bQ\in\calA}}\exp\big\{-n\big(D(Q_1||P_\rmA)+D(Q_2||P_\rmN)+D(Q_3||P_\rmN)\big)\big\}\label{upper:fix}\\
&\le t(M-t)^2(n+1)^{3|\calX|}\max_{\substack{\bQ\in\calA}}\exp\big\{-n\big(D(Q_1||P_\rmA)+D(Q_2||P_\rmN)+D(Q_3||P_\rmN)\big)\big\}\label{number:fix}\\
&\le t(M-t)^2(n+1)^{3|\calX|}\exp\big\{-n\eta(P_\rmA,P_\rmN)\big\}\label{known:eta},
\end{align}
where \eqref{upper:fix} follows from the upper bound on the probability of the type class~\cite[Theorem 11.1.4]{cover2012elements} and $|\calB|=t$, $|\calM_\calB|=M-t$, \eqref{number:fix} follows from the number of types~\cite[Theorem 11.1.1]{cover2012elements} which implies that $|\calP_n(\calX)|\le(n+1)^{|\calX|}$, and \eqref{known:eta} follows from the definition of $\eta(P_1,P_2)$ in \eqref{eta0}.
Analogously, we can obtain the upper bound the probability of $\bar\calE^\mathrm{f,k}_{1,2}$ as follows:
\begin{align}
\bbP_\calB\big\{\bar\calE^\mathrm{f,k}_{1,2}\big\}\le t^2(M-t)(n+1)^{3|\calX|}\exp\big\{-n\eta(P_\rmN,P_\rmA)\big\}.\label{known:eta2}
\end{align}
Combining \eqref{GF1}, \eqref{known:eta} and \eqref{known:eta2}, it follows that
\begin{align}
\beta_\calB(\Phi_\mathrm{fix}|P_\rmN,P_\rmA)\le 4t^2(M-t)^2(n+1)^{3|\calX|}\exp\Big\{-n\min\{\eta(P_\rmA,P_\rmN),\eta(P_\rmN,P_\rmA)\}\Big\}.
\end{align}

Thus, the misclassification exponent satisfies
\begin{align}
-\frac{1}{n}\log\beta_\calB(\Phi_\mathrm{fix}|P_\rmN,P_\rmA)\ge\min\big\{\eta(P_\rmA,P_\rmN),~\eta(P_\rmN,P_\rmA)\big\}.
\end{align}

The proof of Theorem \ref{fixed:known} is now completed.

\subsection{Proof of Theorem \ref{th:known} (Sequential Test with Known Number of Outliers)}
\label{proof:known}
\subsubsection{Expected Stopping Time}

Recall that there are $t$ outliers among $M$ observed sequences. Fix any $\calB\in\calS(t)$ and $n\in\bbN$. The expected stopping time of the sequential test in Algorithm \ref{CA:known} satisfies
\begin{align}
\mathbb{E}_\calB[\tau]=\sum\limits_{k=1}^{\infty}\bbP_\calB\{\tau> k\}=n-1+\sum\limits_{k=n-1}^{\infty}\bbP_\calB\{\tau> k\}.
\end{align}

Recall the sequential test in Algorithm \ref{CA:known}. Fix any $k\in\bbN$. Define the sets $\calC_1$ and $\calC_2$ with respect to the sample size $k$ as $\calC_1^k$ and $\calC_2^k$, respectively. It follows from the test design in step 14 in Algorithm \ref{CA:known} that the sequential test stops if $\min\{|\calC_1|,|\calC_2|\}\ge t$. Thus, the event $\tau>k$ indicates $|\calC_1^k|<t$ or $|\calC_2^k|<t$, which implies that
\begin{align}
\bbP_\calB\{\tau>k\}\le\bbP_\calB\{|\calC_2^k|<t\}+\bbP_\calB\{|\calC_1^k|<t\}.\label{tauk}
\end{align}

Note that in step 4 in Algorithm \ref{CA:known}, the test randomly chooses an index $l\in[M]$ and sets $\hatT_0$ as the type of the sequence $X_l^k$. Define the event $\calW$ such that $\hatT_0$ corresponds to the type of a nominal sample, i.e., $\calW:=\big\{X_l^k\overset{\mathrm{i.i.d.}}\sim P_\rmN\big\}$. Thus, $\calW^c$ denotes the event that $\hatT_0$ is the type of an outlier, i.e., $\calW^c:=\big\{X_l^k\overset{\mathrm{i.i.d.}}\sim P_\rmA\big\}$. The result in \eqref{tauk} can be further upper bounded by
\begin{align}
\nn&\bbP_\calB\{|\calC_2^k|<t\}+\bbP_\calB\{|\calC_1^k|<t\}\\*
&=\bbP_\calB\{|\calC_2^k|<t,~\calW\}+\bbP_\calB\{|\calC_1^k|<t,~\calW\}+\bbP_\calB\{|\calC_2^k|<t,~\calW^c\}+\bbP_\calB\{|\calC_1^k|<t,~\calW^c\}.\label{fourterms}
\end{align}

The first term of \eqref{fourterms} can be upper bounded as follows:
\begin{align}
\nn&\bbP_\calB\{|\calC_2^k|<t,~\calW\}\\*
&\le\bbP_\calB\big\{\exists~i\in\calB,~\mathrm{s.t.}~f\big(\hatT_{X_i^k},\hatT_0\big)\le\lambda_2,~\calW\big\}\label{Bsize}\\
&\le\bbP_\calB\big\{\exists~i\in\calB,~j\in\calM_\calB~\mathrm{s.t.}~f\big(\hatT_{X_i^k},\hatT_{X_j^k}\big)\le\lambda_2\big\}\label{T0}\\
&\le\sum_{i\in\calB}\sum_{j\in\calM_\calB}\bbP_\calB\big\{f\big(\hatT_{X_i^k},\hatT_{X_j^k}\big)\le\lambda_2\big\}\\
&\le \sum_{i\in\calB}\sum_{j\in\calM_\calB}\sum\limits_{\substack{x_i^k,x_j^k\in\calX^{2k}:\\f\big(\hatT_{x_i^k},\hatT_{x_j^k}\big)\le\lambda_2}}
P_\rmA(x^k_i)P_\rmN(x^k_j)\\
&\le \sum_{i\in\calB}\sum_{j\in\calM_\calB}\sum_{\substack{(Q_1,Q_2)\in\calP_k(\calX)^2:\\ f(Q_1,Q_2)\le\lambda_2}}P_\rmA(\calT_{Q_1}^k)\times P_\rmN(\calT_{Q_2}^k)\\
&\le \sum_{i\in\calB}\sum_{j\in\calM_\calB}\sum_{\substack{(Q_1,Q_2)\in\calP_k(\calX)^2:\\ f(Q_1,Q_2)\le\lambda_2}}\exp\big\{-kD(Q_1||P_\rmA)-kD(Q_2||P_\rmN)\big\}\label{upperbound}\\
&\le t(M-t)(k+1)^{2|\calX|}\max_{\substack{(Q_1,Q_2)\in\calP(\calX)^2:\\f(Q_1,Q_2)\le\lambda_2}}\exp\big\{-kD(Q_1||P_\rmA)-kD(Q_2||P_\rmN)\big\}\label{typenumber}\\
&\le t(M-t)\exp\Big\{-k\Big(\Omega(P_\rmA,P_\rmN,\lambda_2)-\frac{2|\calX|\log(k+1)}{k}\Big)\Big\}\\
&\le t(M-t)\exp\Big\{-k\Big(\Omega(P_\rmA,P_\rmN,\lambda_2)-\frac{2|\calX|\log n}{n-1}\Big)\Big\}\label{decrease},
\end{align}
where \eqref{Bsize} follows since $|\calC_2^k|<t$ and $|\calB|=t$ indicate that there exists $i\in\calB$ such that $i\notin\calC_2^k$ and thus $f\big(\hatT_{X_i^k},\hatT_0\big)\le\lambda_2$, \eqref{T0} follows since the event $\calW$ means that $\hatT_0$ corresponds to the type of a nominal sample, \eqref{upperbound} follows from the upper bound on the probability of the type class~\cite[Theorem 11.1.4]{cover2012elements}, \eqref{typenumber} follows from \cite[Theorem 11.1.1]{cover2012elements} which implies that $|\calP_k(\calX)|\le(k+1)^{|\calX|}$ and \eqref{decrease} follows from the fact that $\tfrac{2|\calX|\log k}{k-1}$ is decreasing in $k$ where $k\ge n-1$ and the definition of $\Omega(P_1,P_2,\lambda)$ in \eqref{omega}.

The second term of \eqref{fourterms} satisfies
\begin{align}
\nn&\bbP_\calB\{|\calC_1^k|<t,~\calW\}\\*
&\le\bbP_\calB\big\{\exists~(i,j)\in(\calM_\calB)^2,~i\neq j:f\big(\hatT_{X_i^k},\hatT_{X_j^k}\big)>\lambda_1\big\}\label{C1}\\
&=\sum_{(i,j)\in(\calM_\calB)^2:i\neq j}\sum\limits_{\substack{x_i^k,x_j^k\in\calX^{2k}:\\f\big(\hatT_{X_i^k},\hatT_{X_j^k}\big)>\lambda_1}}
P_\rmN(x^k_i)P_\rmN(x^k_j)\\
&=\sum_{(i,j)\in(\calM_\calB)^2:i\neq j}\sum_{\substack{(Q_1,Q_2)\in\calP_k(\calX)^2:\\ f(Q_1,Q_2)>\lambda_1}}P_\rmN(\calT_{Q_1}^k)\times P_\rmN(\calT_{Q_2}^k)\\
&\le (M-t)^2\max_{\substack{(Q_1,Q_2)\in\calP(\calX)^2:\\f(Q_1,Q_2)>\lambda_1}}\exp\big\{-kD(Q_1||P_\rmN)-kD(Q_2||P_\rmN)+2|\calX|\log(k+1)\big\}\\
&\le (M-t)^2\exp\Big\{-k\Big(\Upsilon(P_\rmN,\lambda_1)-\frac{2|\calX|\log n}{n-1}\Big)\Big\},\label{secondterm}
\end{align}
where \eqref{C1} follows since when $\hatT_0$ corresponds to the type of a nominal sample, $|\calC_1^k|<t$ and $|\calM_\calB|>t$ imply that there exists $i\in\calM_\calB$ such that $i\notin\calC_1^k$ and $f\big(\hatT_{X_i^k},\hatT_0\big)>\lambda_1$, and \eqref{secondterm} follows from the definition of $\Upsilon(P,\lambda)$ in \eqref{upsilon} and the steps analogously to those leading to the result in \eqref{decrease}.

Similarly to \eqref{decrease} and \eqref{secondterm}, we can upper bound the third and fourth terms of \eqref{fourterms} as follows:
\begin{align}
\bbP_\calB\{|\calC_1^k|<t,~\calW^c\}&\le t^2\exp\Big\{-k\Big(\Upsilon(P_\rmA,\lambda_1)-\frac{2|\calX|\log n}{n-1}\Big)\Big\},\label{third}\\
\bbP_\calB\{|\calC_2^k|<t,~\calW^c\}
&\le t(M-t)\exp\Big\{-k\Big(\Omega(P_\rmN,P_\rmA,\lambda_2)-\frac{2|\calX|\log n}{n-1}\Big)\Big\}.\label{fourth}
\end{align}

Combining \eqref{fourterms}, \eqref{decrease}, \eqref{secondterm}, \eqref{third} and \eqref{fourth}, it follows that
\begin{align}
\nn&\sum\limits_{k=n-1}^{\infty}\bbP_\calB\big\{\tau>k\big\}\\*
\nn&\le t^2(M-t)^2\left(\frac{\exp\Big\{-(n-1)\Big(\Omega(P_\rmA,P_\rmN,\lambda_2)-\frac{2|\calX|\log n}{n-1}\Big)\Big\}}{1-\exp\Big\{-\Big(\Omega(P_\rmA,P_\rmN,\lambda_2)-\frac{2|\calX|\log n}{n-1}\Big)\Big\}}+\frac{\exp\Big\{-(n-1)\Big(\Upsilon(P_\rmN,\lambda_1)-\frac{2|\calX|\log n}{n-1}\Big)\Big\}}{1-\exp\Big\{-\Big(\Upsilon(P_\rmN,\lambda_1)-\frac{2|\calX|\log n}{n-1}\Big)\Big\}}\right.\\*
&\qquad\left.+\frac{\exp\Big\{-(n-1)\Big(\Upsilon(P_\rmA,\lambda_1)-\frac{2|\calX|\log n}{n-1}\Big)\Big\}}{1-\exp\Big\{-\Big(\Upsilon(P_\rmA,\lambda_1)-\frac{2|\calX|\log n}{n-1}\Big)\Big\}}+\frac{\exp\Big\{-(n-1)\Big(\Omega(P_\rmN,P_\rmA,\lambda_2)-\frac{2|\calX|\log n}{n-1}\Big)\Big\}}{1-\exp\Big\{-\Big(\Omega(P_\rmN,P_\rmA,\lambda_2)-\frac{2|\calX|\log n}{n-1}\Big)\Big\}}\right)\label{Esk1}\\
&\le 1,\label{known:tau}
\end{align}
when $n$ is sufficiently large and $0<\lambda_1\le\lambda_2<\min\{f(P_\rmA,P_\rmN),~f(P_\rmN,P_\rmA)\}$ since i) $0<\lambda_2<f(P_\rmA,P_\rmN)$ ensures $\Omega(P_\rmA,P_\rmN,\lambda_2)>0$, ii) $\lambda_1>0$ ensures $\Upsilon(P_\rmN,\lambda_1)>0$ and $\Upsilon(P_\rmA,\lambda_1)>0$, and iii) $0<\lambda_2<f(P_\rmN,P_\rmA)$ ensures $\Omega(P_\rmN,P_\rmA,\lambda_1)>0$.

Therefore, under hypothesis $\rmH_\calB$, the expected stopping time of our sequential test in  Algorithm \ref{CA:known} satisfies
\begin{align}
\mathbb{E}_\calB[\tau]= n-1+\sum\limits_{k=n-1}^{\infty}\bbP_\calB\{\tau> k\}\le n,
\end{align}
when $n$ is sufficiently large and $0<\lambda_1\le\lambda_2<\min\{f(P_\rmA,P_\rmN),~f(P_\rmN,P_\rmA)\}$.

\subsubsection{Misclassification Exponent}\label{known:exponent}

Recall our sequential test $\Phi_{\rm{seq}}$ with known number of outliers in Algorithm \ref{CA:known}. Since $|\calC_{\rm out}|=t$, a misclassification event occurs if the following event occurs: $\calE^\mathrm{s,k}$ where in steps 8-12, our test incorrectly claims a nominal sample as an outlier. We consider two cases: $\hatT_0$ chosen randomly in step 4 is the type of a nominal sample or an outlier. When $\hatT_0$ is the type of a nominal sample, $\calE^\mathrm{s,k}$ indicates there exists a nominal sample satisfying $f\big(\hatT_{x_i^\tau},\hatT_0\big)>\lambda_2$. When $\hatT_0$ is the type of an outlier, $\calE^\mathrm{s,k}$ indicates there exists a nominal sample satisfying $f\big(\hatT_{x_i^\tau},\hatT_0\big)<\lambda_1$. Therefore, the probability of the event $\calE^\mathrm{s,k}$ can be upper bounded by the sum of the probabilities of the following two events:
\begin{align}
\bar\calE^\mathrm{s,k}_1&=\Big\{\exists~(i,j)\in(\calM_\calB)^2,~i\neq j:f\big(\hatT_{X_i^\tau},\hatT_{X_j^\tau}\big)>\lambda_2\Big\},\\
\bar\calE^\mathrm{s,k}_2&=\Big\{\exists~i\in\calM_\calB,~j\in\calB:f\big(\hatT_{X_i^\tau},\hatT_{X_j^\tau}\big)<\lambda_1\Big\}.
\end{align}
Furthermore, the probability of the event $\bar\calE^\mathrm{s,k}_1$ can be upper bounded as follows:
\begin{align}
\bbP_\calB\big\{\bar\calE^\mathrm{s,k}_1\big\}
&=\bbP_\calB\big\{\exists~(i,j)\in(\calM_\calB)^2,~i\neq j:f\big(\hatT_{X_i^\tau},\hatT_{X_j^\tau}\big)>\lambda_2\big\}\\
&\le\sum\limits_{k=n-1}^{\infty}\Big(\bbP_\calB\big\{\tau=k\big\}\times\bbP_\calB\big\{\exists~(i,j)\in(\calM_\calB)^2,~i\neq j:f\big(\hatT_{X_i^k},\hatT_{X_j^k}\big)>\lambda_2\big\}\Big)\\
&\le\sum\limits_{k=n-1}^{\infty}\bbP_\calB\big\{\exists~(i,j)\in(\calM_\calB)^2,~i\neq j:f\big(\hatT_{X_i^k},\hatT_{X_j^k}\big)>\lambda_2\big\}\label{prob1}\\
&\le \sum\limits_{k=n-1}^{\infty}(M-t)^2\exp\Big\{-k\Big(\Upsilon(P_\rmN,\lambda_2)-\frac{2|\calX|\log n}{n-1}\Big)\Big\}\label{bsk1}\\
&=(M-t)^2\frac{\exp\Big\{-(n-1)\Big(\Upsilon(P_\rmN,\lambda_2)-\frac{2|\calX|\log n}{n-1}\Big)\Big\}}{1-\exp\Big\{-\Big(\Upsilon(P_\rmN,\lambda_2)-\frac{2|\calX|\log n}{n-1}\Big)\Big\}},\label{sk1}
\end{align}
where \eqref{prob1} follows since $\bbP_\calB\{\tau=k\}\le 1$ and \eqref{bsk1} follows from the steps leading to the results in \eqref{secondterm}.

Analogously to the steps leading to the results in \eqref{sk1}, the probability of $\bar\calE^\mathrm{s,k}_2$ satisfies
\begin{align}
\bbP_\calB\big\{\bar\calE^\mathrm{s,k}_2\big\}
&=\bbP_\calB\big\{\exists~i\in\calM_\calB,~j\in\calB:f\big(\hatT_{X_i^\tau},\hatT_{X_j^\tau}\big)<\lambda_1\big\}\\
&\le t(M-t)\frac{\exp\Big\{-(n-1)\Big(\Omega(P_\rmN,P_\rmA,\lambda_1)-\frac{2|\calX|\log n}{n-1}\Big)\Big\}}{1-\exp\Big\{-\Big(\Omega(P_\rmN,P_\rmA,\lambda_1)-\frac{2|\calX|\log n}{n-1}\Big)\Big\}}.\label{sk2}
\end{align}
Combining \eqref{sk1} and \eqref{sk2} leads to
\begin{align}
\nn&\bbP_\calB\{\calE^\mathrm{s,k}\}\\*
&\le\bbP_\calB\{\bar\calE^\mathrm{s,k}_1\cup\bar\calE^\mathrm{s,k}_2\}\\
&\le A_1\max\left\{\frac{\exp\Big\{-(n-1)\Big(\Upsilon(P_\rmN,\lambda_2)-\frac{2|\calX|\log n}{n-1}\Big)\Big\}}{1-\exp\Big\{-\Big(\Upsilon(P_\rmN,\lambda_2)-\frac{2|\calX|\log n}{n-1}\Big)\Big\}},\frac{\exp\Big\{-(n-1)\Big(\Omega(P_\rmN,P_\rmA,\lambda_1)-\frac{2|\calX|\log n}{n-1}\Big)\Big\}}{1-\exp\Big\{-\Big(\Omega(P_\rmN,P_\rmA,\lambda_1)-\frac{2|\calX|\log n}{n-1}\Big)\Big\}}\right\},\label{sk}
\end{align}
where $A_1:=2t(M-t)^2$.

Thus, the misclassification exponent satisfies
\begin{align}
-\frac{1}{n}\log\beta_\calB(\Phi_\mathrm{seq}|P_\rmN,P_\rmA)&\ge
\min\big\{\Omega(P_\rmN,P_\rmA,\lambda_1),~\Upsilon(P_\rmN,\lambda_2)\big\}.\label{known:mis}
\end{align}

The proof of Theorem \ref{th:known} is now completed.

\subsection{Justification of the Benefit of Sequentiality with Known Number of Outliers}
\label{proof:sequentiality_known}

Recall the definition of $\Omega(P_1,P_2,\lambda)$ in \eqref{omega} and we can rewrite $\Omega(P_\rmA,P_\rmN,\lambda_1)$ as
\begin{align}
\Omega(P_\rmA,P_\rmN,\lambda_1)&=\min_{\substack{(Q_1,Q_2,Q_3)\in\calP(\calX)^3:~f(Q_1,Q_2)\le\lambda_1}}D(Q_1\|P_\rmA)+D(Q_2\|P_\rmN)+D(Q_3\|P_\rmN).\label{com:omega}
\end{align}
Recall the definition of $\eta(P_1,P_2)$ in \eqref{eta0} and we have
\begin{align}
\eta(P_\rmA,P_\rmN)=\min_{\substack{(Q_1,Q_2,Q_3)\in\calP(\calX)^3:f(Q_1,Q_2)\le f(Q_3,Q_2)}}D(Q_1||P_\rmA)+D(Q_2||P_\rmN)+D(Q_3||P_\rmN).\label{etaAN}
\end{align}
Comparing \eqref{com:omega} and \eqref{etaAN}, we obtain $\Omega(P_\rmA,P_\rmN,0)\ge \eta(P_\rmA,P_\rmN)$ since $f(Q_3,Q_2)\ge 0$ for any pair of distributions $(Q_2,Q_3)\in\calP(\calX)^2$.

Furthermore, by letting $Q_1=P_\rmA$, it follows from \eqref{etaAN} that
\begin{align}
\eta(P_\rmA,P_\rmN)&\le\min_{\substack{(Q_2,Q_3)\in\calP(\calX)^2:f(Q_3,Q_2)\ge f(P_\rmA,Q_2)}}D(Q_2||P_\rmN)+D(Q_3||P_\rmN)\\
&=\min_{\substack{(Q_1,Q_2)\in\calP(\calX)^2:f(Q_1,Q_2)\ge f(P_\rmA,Q_2)}}D(Q_1||P_\rmN)+D(Q_2||P_\rmN).\label{etale}
\end{align}
Define the feasible region of $\Upsilon(P_\rmN,f(P_\rmA,P_\rmN))$ in \eqref{upsilon} and the right hand side of \eqref{etale} as $\calF_\Upsilon:=\{(Q_1,Q_2)\in\calP(\calX)^2:~f(Q_1,$ $Q_2)$ $\ge f(P_\rmA,P_\rmN)\}$ and $\calF_\eta:=\{(Q_1,Q_2)\in\calP(\calX)^2:~f(Q_1,Q_2)\ge f(P_\rmA,Q_2)\}$. Furthermore, define
\begin{align}
\lambda^*(P_\rmA)=\min_{(Q_1,Q_2)\in\calF_\eta}f(P_\rmA,Q_2).\label{lambdastar}
\end{align}
To show $\calF_\Upsilon\subseteq\calF_\eta$, it suffices to prove $\lambda^*(P_\rmA)\le f(P_\rmA,P_\rmN)$. By letting $(Q_1,Q_2)=(P_\rmA,P_\rmN)$ which satisfies the constraint function of \eqref{lambdastar}, the objective function of \eqref{lambdastar} is $f(P_\rmA,P_\rmN)$ and thus, we obtain $\lambda^*(P_\rmA)\le f(P_\rmA,P_\rmN)$. Subsequently, it follows that $\Upsilon(P_\rmN,f(P_\rmA,P_\rmN))\ge \eta(P_\rmA,P_\rmN)$.

Therefore, we conclude that
\begin{align}\label{benefit:known}
\min\big\{\eta(P_\rmA,P_\rmN),~\eta(P_\rmN,P_\rmA)\big\}\le\eta(P_\rmA,P_\rmN)\le\min\{\Omega(P_\rmA,P_\rmN,0),\Upsilon(P_\rmN,f(P_\rmA,P_\rmN))\}.
\end{align}

\subsection{Proof of Theorem \ref{fixed:unknown} (Fixed-length Test with Unknown Number of Outliers)}
\label{proof:fix_unknown}
When the number of outliers is unknown, the theoretical benchmark is the exponents for misclassification, false reject and false alarm probabilities in \eqref{def:mis}, \eqref{def:fr} and \eqref{def:fa}, respectively. Recall our fixed-length test in  Algorithm \ref{FA:unknown}.

\subsubsection{False Alarm Probability}
Recall the definition of $\calM_{\rm{dis}}$. The false alarm probability of the test in Algorithm \ref{FA:unknown} satisfies
\begin{align}
\mathrm{P_{fa}}(\Phi|P_\rmA,P_\rmN)
&=\bbP_\rmr\{\Phi(\bX^n)\ne\rmH_\rmr\}\\
&=\bbP_\rmr\Big\{\max_{(i,j)\in\calM_{\rm{dis}}}f\big(\hatT_{X_i^n},\hatT_{X_j^n}\big)>\lambda\Big\}\label{fix:faprob}\\
&=\bbP_\rmr\big\{\exists~(i,j)\in\calM_{\rm{dis}},~{\rm s.t.}~f\big(\hatT_{X_i^n},\hatT_{X_j^n}\big)>\lambda\big\}\\
&\le\sum\limits_{(i,j)\in\calM_{\rm{dis}}}\bbP_\rmr\big\{f\big(\hatT_{X_i^n},\hatT_{X_j^n}\big)>\lambda\big\}\\
&=\sum\limits_{(i,j)\in\calM_{\rm{dis}}}\sum\limits_{\substack{x_i^n,x_j^n\in\calX^{2n}:\\f\big(\hatT_{x_i^n},\hatT_{x_j^n}\big)>\lambda}}
P_\rmN(x^n_i)P_\rmN(x^n_j)\\
&=\sum\limits_{(i,j)\in\calM_{\rm{dis}}}\sum\limits_{\substack{(Q_1,Q_2)\in\calP_n(\calX):\\f(Q_1,Q_2)>\lambda}}
P_\rmN(\calT_{Q_1}^n)P_\rmN(\calT_{Q_2}^n)\\
&\le \sum\limits_{(i,j)\in\calM_{\rm{dis}}}(n+1)^{2|\calX|}\max_{\substack{(Q_1,Q_2)\in\calP_n(\calX):\\f(Q_1,Q_2)>\lambda}}
\exp\big\{-n\big(D(Q_1||P_\rmN)+D(Q_2||P_\rmN)\big)\big\}\label{upper:fix_unknown}\\
&\le M(M-1)\exp\Big\{-n\Big(\Upsilon(P_\rmN,\lambda)-\frac{2|\calX|\log(n+1)}{n}\Big)\Big\},\label{unknown:fix_upsilon}
\end{align}
where \eqref{fix:faprob} follows from step 2 of outlier detection phase in Algorithm \ref{FA:unknown}, \eqref{upper:fix_unknown} follows from the upper bound on the probability of the type class~\cite[Theorem 11.1.4]{cover2012elements} and the upper bound on the number of types which implies that $|\calP_n(\calX)|\le(n+1)^{|\calX|}$~\cite[Theorem 11.1.1]{cover2012elements} and \eqref{unknown:fix_upsilon} follows from the definition of $\Upsilon(P,\lambda)$ in \eqref{upsilon}.

Thus, the false alarm exponent satisfies
\begin{align}
-\frac{1}{n}\log \mathrm{P_{fa}}(\Phi_\mathrm{fix}^{\rmu}|P_\rmN,P_\rmA)&\ge\Upsilon(P_\rmN,\lambda).
\end{align}

\subsubsection{False Reject Probability}
Fix any $\calB\in\calS$. Under hypothesis $\rmH_\calB$, the false reject probability satisfies
\begin{align}
\zeta_\calB(\Phi|P_\rmA,P_\rmN)
&=\bbP_\calB\{\Phi(\bX^n)=\rmH_\rmr\}\\
&=\bbP_\calB\Big\{\max_{(i,j)\in\calM_{\rm{dis}}}f\big(\hatT_{X_i^n},\hatT_{X_j^n}\big)\le\lambda\Big\}\label{fix:frprob}\\
&=\bbP_\calB\Big\{\forall~(i,j)\in\calM_{\rm{dis}}~\mathrm{s.t.}~f\big(\hatT_{X_i^n},\hatT_{X_j^n}\big)\le\lambda\Big\}\\
\nn&\le\min\Big\{\bbP_\calB\big\{\forall~i\in\calB,~j\in\calM_\calB~\mathrm{s.t.}~f\big(\hatT_{X_i^n},\hatT_{X_j^n}\big)\le\lambda\big\},\\*
&\qquad\qquad\bbP_\calB\big\{\forall~j\in\calB,~i\in\calM_\calB~\mathrm{s.t.}~f\big(\hatT_{X_i^n},\hatT_{X_j^n}\big)\le\lambda\big\}\Big\},\label{fr:twoterms}
\end{align}
where \eqref{fix:frprob} follows from step 2 of outlier detection in Algorithm \ref{FA:unknown}.

The first term of \eqref{fr:twoterms} can be upper bounded as follows:
\begin{align}
\nn&\bbP_\calB\big\{\forall~i\in\calB,~j\in\calM_\calB~\mathrm{s.t.}~f\big(\hatT_{X_i^n},\hatT_{X_j^n}\big)\le\lambda\big\}\\*
&\le\max_{i\in\calB,j\in\calM_\calB}\sum\limits_{\substack{x_i^n,x_j^n\in\calX^{2n}:\\f\big(\hatT_{x_i^n},\hatT_{x_j^n}\big)\le\lambda}}P_\rmA(x^n_i)P_\rmN(x^n_j)\\
&=\max_{i\in\calB,j\in\calM_\calB}\sum\limits_{\substack{(Q_1,Q_2)\in\calP_n(\calX):\\f(Q_1,Q_2)\le\lambda}}
P_\rmA(\calT_{Q_1}^n)P_\rmN(\calT_{Q_2}^n)\\
&\le(n+1)^{2|\calX|}
\max_{\substack{(Q_1,Q_2)\in\calP_n(\calX)^2:\\f(Q_1,Q_2)\le\lambda}}\exp\big\{-n\big(D(Q_1||P_\rmA)+D(Q_2||P_\rmN)\big)\big\}\\
&\le\exp\Big\{-n\Big(\Omega(P_\rmA,P_\rmN,\lambda)-\frac{2|\calX|\log(n+1)}{n}\Big)\Big\},\label{fix:upper_fr}
\end{align}
where \eqref{fix:upper_fr} follows from the steps analogously to those leading to the result in \eqref{unknown:fix_upsilon} and the definition of $\Omega(P_1,P_2,\lambda)$ in \eqref{omega}. Similarly, the second term of \eqref{fr:twoterms} can be upper bounded as
\begin{align}
\bbP_\calB\big\{\forall~j\in\calB,~i\in\calM_\calB~\mathrm{s.t.}~f\big(\hatT_{X_i^n},\hatT_{X_j^n}\big)\le\lambda\big\}
\le\exp\Big\{-n\Big(\Omega(P_\rmN,P_\rmA,\lambda)-\frac{2|\calX|\log(n+1)}{n}\Big)\Big\}.\label{fr:second}
\end{align}

Thus, combining \eqref{fr:twoterms}, \eqref{fix:upper_fr} and \eqref{fr:second}, the false reject exponent satisfies
\begin{align}
-\frac{1}{n}\zeta_\calB(\Phi_\mathrm{fix}^{\rmu}|P_\rmN,P_\rmA)&\ge\max\{\Omega(P_\rmA,P_\rmN,\lambda),~\Omega(P_\rmN,P_\rmA,\lambda)\}.
\end{align}

\subsubsection{Misclassification Probability}
A misclassification event of the test in Algorithm \ref{FA:unknown} occurs if one of the following two error events occurs: i) $\calE^\mathrm{f,u}_1$ where in step 7, the two cluster centers are types of either two outliers or two nominal samples, and ii) $\calE^\mathrm{f,u}_2$ where in steps 11-12, an outlier is incorrectly identified as a nominal sample or a nominal sample is incorrectly classified as an outlier when $(\calE^\mathrm{f,u}_1)^\rmc$ occurs.
Thus, it follows that
\begin{align}
\beta_\calB(\Phi|P_\rmN,P_\rmA)
&=\bbP_\calB\{\Phi(\bX^n)\notin\{\rmH_\rmr, \rmH_\calB\}\}\\
&\le\bbP_\calB\{\calE^\mathrm{f,u}_1\}+\bbP_\calB\{\calE^\mathrm{f,u}_2\}.
\end{align}

The error event $\calE^\mathrm{f,u}_1$ can be further categorized into two events: $\calE^\mathrm{f,u}_{1,1}$ when both cluster centers are types of nominal samples and $\calE^\mathrm{f,u}_{1,2}$ when both cluster centers are types of outliers.
We first analyze the error event $\calE^\mathrm{f,u}_{1,1}$ when both cluster centers are types of nominal samples. Let $\hatT_0$ be $\hatT_{x_l^n}$ chosen randomly in step 5 of Algorithm \ref{FA:unknown}. In this case, $\hatT_0$ corresponds to the type of a nominal sample. It follows from steps 5-7 of Algorithm \ref{FA:unknown} that the event $(\calE^\mathrm{f,u}_{1,1})^\rmc$ occurs if the scoring function between the type of any outlier and $\hatT_0$ is greater than the scoring function between the type of any nominal sample and $\hatT_0$.
Thus, the event $\calE^\mathrm{f,u}_{1,1}$ implies there exists an outlier and a nominal sample such that the scoring function between the type of the outlier and $\hatT_0$ is smaller than the scoring function between the type of the nominal sample and $\hatT_0$, which is exactly the event $\bar\calE^\mathrm{f,k}_{1,1}$ (cf. \eqref{def:calR1'}). Analogously, the probability of the event $\calE^\mathrm{f,u}_{1,2}$ can be upper bounded by the probability of the event $\bar\calE^\mathrm{f,k}_{1,2}$ (cf. \eqref{def:calR2'}).
Thus, following from the steps leading to the result in \eqref{known:eta}, the probability of the event $\calE^\mathrm{f,u}_1$ can be upper bounded as follows:
\begin{align}
\bbP_\calB\{\calE^\mathrm{f,u}_1\}&=\bbP_\calB\{\calE^\mathrm{f,u}_{1,1}\}+\bbP_\calB\{\calE^\mathrm{f,u}_{1,1}\}\\
&\le\bbP_\calB\{\bar\calE^\mathrm{f,k}_{1,1}\}+\bbP_\calB\{\bar\calE^\mathrm{f,k}_{1,2}\}\\
&\le 2t^2(M-t)^2\exp\big\{-n\min\big\{\eta(P_\rmA,P_\rmN),~\eta(P_\rmN,P_\rmA)\big\}+3|\calX|\log(n+1)\big\}.\label{mis1}
\end{align}

The error event $\calE^\mathrm{f,u}_2$ can also be categorized into two events: $\calE^\mathrm{f,u}_{2,1}$ where an outlier is incorrectly identified as a nominal sample and $\calE^\mathrm{f,u}_{2,2}$ where a nominal sample is incorrectly classified as an outlier, when the two cluster centers $c_1$ and $c_2$ are types of a nominal sample and an outlier. Without loss of generality, let $c_1$ correspond to the type of a nominal sample and $c_2$ correspond to the type of an outlier. It follows that
\begin{align}
\calE^\mathrm{f,u}_{2,1}&:=(\calE^\mathrm{f,u}_1)^\rmc\cap
\Big\{\exists~i\in\calB:f\Big(\hatT_{X_i^n},c_1\Big)\le f\Big(\hatT_{X_i^n},c_2\Big)\Big\},\\
\calE^\mathrm{f,u}_{2,2}&:=(\calE^\mathrm{f,u}_1)^\rmc\cap
\Big\{\exists~j\in\calM_\calB:f\Big(\hatT_{X_j^n},c_2\Big)\le f\Big(\hatT_{X_j^n},c_1\Big)\Big\}.
\end{align}
Since $c_1$ is the type of a nominal sample whose index belongs to the set $\calM_\calB$ and $c_2$ is the type of an outlier whose index belongs to the set $\calB$, the probability of the events $\calE^\mathrm{f,u}_{2,1}$ and $\calE^\mathrm{f,u}_{2,2}$ can be upper bounded by the probability of the following events:
\begin{align}
\bar\calE^\mathrm{f,u}_{2,1}&:=(\calE^\mathrm{f,u}_1)^\rmc\cap\Big\{\exists~(i_1,i_2)\in\calB^2,i_1\neq i_2,~\exists~j\in\calM_\calB:~
f\Big(\hatT_{X_{i_1}^n},\hatT_{X_j^n}\Big)\le f\Big(\hatT_{X_{i_1}^n},\hatT_{X_{i_2}^n}\Big)\Big\},\\
\bar\calE^\mathrm{f,u}_{2,2}&:=(\calE^\mathrm{f,u}_1)^\rmc\cap\Big\{\exists~i\in\calB,~\exists~(j_1,j_2)\in(\calM_\calB)^2,~j_1\neq j_2:~
f\Big(\hatT_{X_{j_1}^n},\hatT_{X_i^n}\Big)\le f\Big(\hatT_{X_{j_1}^n},\hatT_{X_{j_2}^n}\Big)\Big\}.
\end{align}

Define the set
\begin{align}
\calC=\{(Q_1,Q_2,Q_3)\in\calP(\calX)^3:f(Q_1,Q_3)\le f(Q_1,Q_2)\}.
\end{align}
Analogously to the steps leading to the result in \eqref{known:eta}, the probability of the event $\bar\calE^\mathrm{f,u}_{2,1}$ can be upper bounded as follows:
\begin{align}
\bbP_\calB\{\bar\calE^\mathrm{f,u}_{2,1}\}
&\le\bbP_\calB\Big\{\exists~(i_1,i_2)\in\calB^2,i_1\neq i_2,~\exists~j\in\calM_\calB:~f\Big(\hatT_{X_{i_1}^n},\hatT_{X_j^n}\Big)<f\Big(\hatT_{X_{i_1}^n},\hatT_{X_{i_2}^n}\Big)\Big\}\\
&\le\sum_{\substack{(i_1,i_2)\in\calB^2,\\i_1\neq i_2}}\sum_{j\in\calM_\calB}\bbP_\calB\Big\{f\Big(\hatT_{X_{i_1}^n},\hatT_{X_j^n}\Big)<f\Big(\hatT_{X_{i_1}^n},\hatT_{X_{i_2}^n}\Big)\Big\}\\
&\le \sum_{\substack{(i_1,i_2)\in\calB^2,\\i_1\neq i_2}}\sum_{j\in\calM_\calB}\sum\limits_{\substack{\bQ\in\calC}}P_\rmA(\calT_{Q_1}^n)P_\rmA(\calT_{Q_2}^n)P_\rmN(\calT_{Q_3}^n)\\
&\le t^2(M-t)(n+1)^{3|\calX|}\max_{\substack{\bQ\in\calC}}\exp\big\{-n\big(D(Q_1||P_\rmA)+D(Q_2||P_\rmA)+D(Q_3||P_\rmN)\big)\big\}\\
&\le t^2(M-t)\exp\big\{-n\gamma(P_\rmA,P_\rmN)+3|\calX|\log(n+1)\big\},\label{delta:unknown1}
\end{align}
where \eqref{delta:unknown1} follows from the definition of $\gamma(P_1,P_2)$ in \eqref{gamma}.
Similarly, it follows that
\begin{align}
\bbP_\calB\{\bar\calE^\mathrm{f,u}_{2,2}\}\le t(M-t)^2\exp\big\{-n \gamma(P_\rmN,P_\rmA)+3|\calX|\log(n+1)\big\}.
\label{delta:unknown2}
\end{align}

Combining \eqref{mis1}, \eqref{delta:unknown1} and \eqref{delta:unknown2}, the misclassification exponent satisfies
\begin{align}
-\frac{1}{n}\beta_\calB(\Phi_\mathrm{fix}^{\rmu}|P_\rmN,P_\rmA)\ge\min\big\{\eta(P_\rmN,P_\rmA),~\eta(P_\rmA,P_\rmN),~\gamma(P_\rmA,P_\rmN),~\gamma(P_\rmN,P_\rmA)\big\}.
\end{align}

\subsection{Proof of Theorem \ref{th:unknown} (Sequential Test with Unknown Number of Outliers)}
\label{proof:unknown}

When the number of outliers is unknown, the theoretical benchmark is the exponents for misclassification, false reject and false alarm probabilities in \eqref{def:mis}, \eqref{def:fr} and \eqref{def:fa}, respectively. Recall our sequential test $\Phi_\mathrm{seq}^{\rmu}$ in  Algorithm \ref{CA:unknown}. We first consider the null hypothesis, show that our test satisfies the expected stopping time universality constraint under mild conditions and bound the achievable false alarm exponent. Subsequently, we consider each non-null hypothesis, and bound the achievable false reject and misclassification exponents.

\subsubsection{Analysis under Null Hypothesis}

We first prove our test $\Phi_\mathrm{seq}^{\rmu}$ satisfies expected stopping time universality constraint under the null hypothesis.
The average stopping time under hypothesis $\rmH_\rmr$ can be expressed as the following form:
\begin{align}\label{Ertau:unknown}
\mathbb{E}_\rmr[\tau]=\sum\limits_{k=1}^{\infty}\bbP_\rmr\{\tau> k\}=n-1+\sum\limits_{k=n-1}^{\infty}\bbP_\rmr\{\tau> k\}.
\end{align}

The second term of \eqref{Ertau:unknown} satisfies
\begin{align}
\sum\limits_{k=n-1}^{\infty}\bbP_\rmr\{\tau>k\}
&\le\sum\limits_{k=n-1}^{\infty}\bbP_\rmr\big\{\lambda_1<\max_{(i,j)\in\calM_{\rm{dis}}}f\big(\hatT_{X_i^k},\hatT_{X_j^k}\big)\le\lambda_2\big\}\label{tau4}\\
&\le\sum\limits_{k=n-1}^{\infty}\bbP_\rmr\big\{\max_{(i,j)\in\calM_{\rm{dis}}}f\big(\hatT_{X_i^k},\hatT_{X_j^k}\big)>\lambda_1\big\}\\
&\le\sum\limits_{k=n-1}^{\infty}\bbP_\rmr\big\{\exists~(i,j)\in\calM_{\rm{dis}},~{\rm s.t.}~f\big(\hatT_{X_i^k},\hatT_{X_j^k}\big)>\lambda_1\big\}\\
&\le\sum\limits_{k=n-1}^{\infty}\sum\limits_{(i,j)\in\calM_{\rm{dis}}}\bbP_\rmr\big\{f\big(\hatT_{X_i^k},\hatT_{X_j^k}\big)>\lambda_1\big\}\\
&\le \sum\limits_{k=n-1}^{\infty}\sum\limits_{(i,j)\in\calM_{\rm{dis}}}\sum\limits_{\substack{x_i^k,x_j^k\in\calX^{2k}:\\f\big(\hatT_{x_i^k},\hatT_{x_j^k}\big)>\lambda_1}}
P_\rmN(x^k_i)P_\rmN(x^k_j)\\
&\le \sum\limits_{k=n-1}^{\infty}\sum\limits_{(i,j)\in\calM_{\rm{dis}}}\sum\limits_{\substack{(Q_1,Q_2)\in\calP_k(\calX):\\f(Q_1,Q_2)>\lambda_1}}
P_\rmN(\calT_{Q_1}^k)P_\rmN(\calT_{Q_2}^k)\\
&\le \sum\limits_{k=n-1}^{\infty}M(M-1)\max_{\substack{(Q_1,Q_2)\in\calP_k(\calX):\\f(Q_1,Q_2)>\lambda_1}}
\exp\big\{-k\big(D(Q_1||P_\rmN)+D(Q_2||P_\rmN)\big)+2|\calX|\log(k+1)\big\}\label{upperbound2}\\
&\le\sum\limits_{k=n-1}^{\infty}M(M-1)\exp\Big\{-k\Big(\Upsilon(P_\rmN,\lambda_1)-\frac{2|\calX|\log(k+1)}{k}\Big)\Big\}\label{unknown:upsilon}\\
&\le\sum\limits_{k=n-1}^{\infty}M(M-1)\exp\Big\{-k\Big(\Upsilon(P_\rmN,\lambda_1)-\frac{2|\calX|\log n}{n-1}\Big)\Big\}\label{unknown:decrease}\\
&=M(M-1)\frac{\exp\Big\{-(n-1)\Big(\Upsilon(P_\rmN,\lambda_1)-\frac{2|\calX|\log n}{n-1}\Big)\Big\}}{1-\exp\Big\{-\Big(\Upsilon(P_\rmN,\lambda_1)-\frac{2|\calX|\log n}{n-1}\Big)\Big\}}\label{unknown:sumr}\\
&\le 1\label{unknown:taur2},
\end{align}
where \eqref{tau4} follows from the definition of the stopping time in Algorithm \ref{CA:unknown}, \eqref{upperbound2} follows from the upper bound on the probability of the type class~\cite[Theorem 11.1.4]{cover2012elements} and the upper bound on the number of types which implies that $|\calP_k(\calX)|\le(k+1)^{|\calX|}$~\cite[Theorem 11.1.1]{cover2012elements}, \eqref{unknown:upsilon} follows from the definition of $\Upsilon(P,\lambda)$ in \eqref{upsilon}, \eqref{unknown:decrease} follows since $\tfrac{2|\calX|\log k}{k-1}$ is decreasing in $k$ and \eqref{unknown:taur2} holds when $n$ is sufficiently large and $\lambda_1>0$.

Therefore, under hypothesis $\rmH_\rmr$, the expected stopping time of our sequential test in  Algorithm \ref{CA:unknown} satisfies
\begin{align}
\mathbb{E}_\rmr[\tau]= n-1+\sum\limits_{k=n-1}^{\infty}\bbP_\rmr\{\tau> k\}\le n,
\end{align}
when $n$ is sufficiently large and $\lambda_1>0$.

The false alarm probability satisfies
\begin{align}
\mathrm{P_{fa}}(\Phi|P_\rmA,P_\rmN)
&=\bbP_\rmr\{\Phi(\bX^\tau)\ne\rmH_\rmr\}\\
&=\bbP_\rmr\Big\{\max_{(i,j)\in\calM_{\rm{dis}}}f\big(\hatT_{x_i^\tau},\hatT_{x_j^\tau}\big)>\lambda_2\Big\}\label{faprob}\\
&\le M(M-1)
\frac{\exp\Big\{-(n-1)\Big(\Upsilon(P_\rmN,\lambda_2)-\frac{2|\calX|\log n}{n-1}\Big)\Big\}}{1-\exp\Big\{-\Big(\Upsilon(P_\rmN,\lambda_2)-\frac{2|\calX|\log n}{n-1}\Big)\Big\}}.\label{fa:sumk}
\end{align}
where \eqref{faprob} follows from step 5 of outlier detection phase in Algorithm \ref{CA:unknown}, and \eqref{fa:sumk} follows from the steps analogously to those leading to the result in \eqref{unknown:sumr}.

Thus, the false alarm exponent satisfies
\begin{align}
-\frac{1}{n}\log \mathrm{P_{fa}}(\Phi_\mathrm{seq}^{\rmu}|P_\rmN,P_\rmA)&\ge\Upsilon(P_\rmN,\lambda_2).
\end{align}

\subsubsection{Analysis under Non-Null Hypotheses}

Fix any $\calB\in\calS$. We now prove our test $\Phi_\mathrm{seq}^{\rmu}$ satisfies expected stopping time universality constraint under the non-null hypothesis $\rmH_\calB$.
Similarly to \eqref{Ertau:unknown}, the average stopping time under hypothesis $\rmH_\calB$ satisfies
\begin{align}\label{Etau:unknown}
\mathbb{E}_\calB[\tau]=n-1+\sum\limits_{k=n-1}^{\infty}\bbP_\calB\{\tau> k\}.
\end{align}

Analogously to the steps leading to the result in \eqref{unknown:sumr}, the second term of \eqref{Etau:unknown} satisfies
\begin{align}
\nn&\sum\limits_{k=n-1}^{\infty}\bbP_\calB\{\tau>k\}\\*
&\le\sum\limits_{k=n-1}^{\infty}\bbP_\calB\Big\{\lambda_1<\max_{(i,j)\in\calM_{\rm{dis}}}f\big(\hatT_{x_i^k},\hatT_{x_j^k}\big)\le\lambda_2\Big\}\\
&\le\sum\limits_{k=n-1}^{\infty}\bbP_\calB\Big\{\forall~(i,j)\in\calM_{\rm{dis}}~\mathrm{s.t.}~f\big(\hatT_{x_i^k},\hatT_{x_j^k}\big)\le\lambda_2\Big\}\\
&\le\sum\limits_{k=n-1}^{\infty}\min\Big\{\bbP_\calB\big\{\forall~i\in\calB,~j\in\calM_\calB~\mathrm{s.t.}~f\big(\hatT_{x_i^k},\hatT_{x_j^k}\big)\le\lambda_2\big\},~
\bbP_\calB\big\{\forall~j\in\calB,~i\in\calM_\calB~\mathrm{s.t.}~f\big(\hatT_{x_i^k},\hatT_{x_j^k}\big)\le\lambda_2\big\}\Big\}\label{unknown:tau}\\
\nn&\le\sum\limits_{k=n-1}^{\infty}\min\bigg\{\exp\Big\{-(n-1)\Big(\Omega(P_\rmA,P_\rmN,\lambda_2)-\frac{2|\calX|\log n}{n-1}\Big)\Big\},\\*
&\qquad\qquad\qquad\:\:\exp\Big\{-(n-1)\Big(\Omega(P_\rmN,P_\rmA,\lambda_2)-\frac{2|\calX|\log n}{n-1}\Big)\Big\}\bigg\}\\
&\le\min\left\{\frac{\exp\Big\{-(n-1)\Big(\Omega(P_\rmA,P_\rmN,\lambda_2)-\frac{2|\calX|\log n}{n-1}\Big)\Big\}}{1-\exp\Big\{-\Big(\Omega(P_\rmA,P_\rmN,\lambda_2)-\frac{2|\calX|\log n}{n-1}\Big)\Big\}},~\frac{\exp\Big\{-(n-1)\Big(\Omega(P_\rmN,P_\rmA,\lambda_2)-\frac{2|\calX|\log n}{n-1}\Big)\Big\}}{1-\exp\Big\{-\Big(\Omega(P_\rmN,P_\rmA,\lambda_2)-\frac{2|\calX|\log n}{n-1}\Big)\Big\}}\right\}\label{unknown:sum}\\
&\le 1\label{unknown:tau2},
\end{align}
where \eqref{unknown:tau2} holds when $n$ is sufficiently large and $0<\lambda_2<\min\{f(P_\rmA,P_\rmN),~f(P_\rmN,P_\rmA)\}$ since $0<\lambda_2<f(P_\rmA,P_\rmN)$ ensures $\Omega(P_\rmA,P_\rmN,\lambda_2)>0$ and $0<\lambda_2<f(P_\rmN,P_\rmA)$ ensures $\Omega(P_\rmN,P_\rmA,\lambda_2)>0$.

Therefore, under hypothesis $\rmH_\calB$, the expected stopping time of our sequential test in Algorithm \ref{CA:unknown} satisfies
\begin{align}
\mathbb{E}_\calB[\tau]= n-1+\sum\limits_{k=n-1}^{\infty}\bbP_\calB\{\tau> k\}\le n,
\end{align}
when $n$ is sufficiently large and $0<\lambda_2<\min\{f(P_\rmA,P_\rmN),~f(P_\rmN,P_\rmA)\}$.

Furthermore, under hypothesis $\rmH_\calB$, the false reject probability satisfies
\begin{align}
\nn&\zeta_\calB(\Phi|P_\rmA,P_\rmN)\\
&=\bbP_\calB\{\Phi(\bX^\tau)=\rmH_\rmr\}\\
&=\bbP_\calB\Big\{\max_{(i,j)\in\calM_{\rm{dis}}}f\big(\hatT_{X_i^\tau},\hatT_{X_j^\tau}\big)\le\lambda_1\Big\}\label{frprob}\\
&\le\sum\limits_{k=n-1}^{\infty}\bbP_\calB\Big\{\tau=k,\max_{(i,j)\in\calM_{\rm{dis}}}f\big(\hatT_{X_i^k},\hatT_{X_j^k}\big)\le\lambda_1\Big\}\\
&\le\sum\limits_{k=n-1}^{\infty}\bbP_\calB\Big\{\max_{(i,j)\in\calM_{\rm{dis}}}f\big(\hatT_{X_i^k},\hatT_{X_j^k}\big)\le\lambda_1\Big\}\\
&\le\sum\limits_{k=n-1}^{\infty}\bbP_\calB\Big\{\forall~(i,j)\in\calM_{\rm{dis}}~\mathrm{s.t.}~f\big(\hatT_{x_i^k},\hatT_{x_j^k}\big)\le\lambda_1\Big\}\\
&\le \min\left\{\frac{\exp\Big\{-(n-1)\Big(\Omega(P_\rmA,P_\rmN,\lambda_1)-\frac{2|\calX|\log n}{n-1}\Big)\Big\}}{1-\exp\Big\{-\Big(\Omega(P_\rmA,P_\rmN,\lambda_1)-\frac{2|\calX|\log n}{n-1}\Big)\Big\}},\frac{\exp\Big\{-(n-1)\Big(\Omega(P_\rmN,P_\rmA,\lambda_1)-\frac{2|\calX|\log n}{n-1}\Big)\Big\}}{1-\exp\Big\{-\Big(\Omega(P_\rmN,P_\rmA,\lambda_1)-\frac{2|\calX|\log n}{n-1}\Big)\Big\}}\right\},\label{fr:sumk}
\end{align}
where \eqref{frprob} follows from step 6 of outlier detection phase in Algorithm \ref{CA:unknown}, and \eqref{fr:sumk} follows from the steps analogously to those leading to the result in \eqref{unknown:sum}.

Thus, the false reject exponent satisfies
\begin{align}
-\frac{1}{n}\zeta_\calB(\Phi_\mathrm{seq}^{\rmu}|P_\rmN,P_\rmA)&\ge\max\{\Omega(P_\rmA,P_\rmN,\lambda_1),~\Omega(P_\rmN,P_\rmA,\lambda_1)\}.
\end{align}

Finally, we analyze the misclassification probability under hypothesis $\rmH_\calB$.
A misclassification error event occurs if one of the following two error events occurs: $\calE^\mathrm{s,u}_1$ where a nominal sample is falsely identified as an outlier and $\calE^\mathrm{s,u}_2$ where an outlier is falsely identified as a nominal sample. Thus, it follows that
\begin{align}
\beta_\calB(\Phi|P_\rmN,P_\rmA)&=\bbP_\calB\{\Phi(\bX^\tau)\notin\{\rmH_\rmr, \rmH_\calB\}\}\\
&\le\bbP_\calB\{\calE^\mathrm{s,u}_1\}+\bbP_\calB\{\calE^\mathrm{s,u}_2\}\label{su}.
\end{align}
Note that the first error event $\calE^\mathrm{s,u}_1$ is equivalent to the error event $\calE^\mathrm{s,k}$ for our sequential test with known number of outliers in Algorithm \ref{CA:known}, which was analyzed in \eqref{sk}.

We next consider the second error event $\calE^\mathrm{s,u}_2$ that our test claims an outlier as a nominal sample. When $\hatT_0$ is the type of a nominal sample, $\calE^\mathrm{s,u}_2$ indicates there exists an outlier satisfying $f\big(\hatT_{x_i^\tau},\hatT_0\big)<\lambda_1$. When $\hatT_0$ is the type of an outlier, $\calE^\mathrm{s,u}_2$ indicates there exists an outlier satisfying $f\big(\hatT_{x_i^\tau},\hatT_0\big)>\lambda_2$. Therefore, the probability of the event $\calE^\mathrm{s,u}_2$ can be upper bounded by the probability of the following event $\bar\calE^\mathrm{s,u}_2$:
\begin{align}
\bar\calE^\mathrm{s,u}_2:&=\Big\{\exists~i\in\calB,~j\in\calM_\calB:f\big(\hatT_{X_i^\tau},\hatT_{X_j^\tau}\big)<\lambda_1\Big\}\bigcup \Big\{\exists~(i,j)\in\calB^2,~i\neq j:f\big(\hatT_{X_i^\tau},\hatT_{X_j^\tau}\big)>\lambda_2\Big\}.
\end{align}

Analogously to the steps leading to the result in \eqref{sk}, $\bar\calE^\mathrm{s,u}_2$ can be upper bounded as follows:
\begin{align}
\nn&\bbP_\calB\{\bar\calE^\mathrm{s,u}_2\}\\*
&\le A_2\frac{\exp\Big\{-(n-1)\Big(\Omega(P_\rmA,P_\rmN,\lambda_1)-\frac{2|\calX|\log n}{n-1}\Big)\Big\}}{1-\exp\Big\{-\Big(\Omega(P_\rmA,P_\rmN,\lambda_1)-\frac{2|\calX|\log n}{n-1}\Big)\Big\}}
+|\calB|^2\frac{\exp\Big\{-(n-1)\Big(\Upsilon(P_\rmA,\lambda_2)-\frac{2|\calX|\log n}{n-1}\Big)\Big\}}{1-\exp\Big\{-\Big(\Upsilon(P_\rmA,\lambda_2)-\frac{2|\calX|\log n}{n-1}\Big)\Big\}},\label{su:second}
\end{align}
where $A_2:=|\calB|(M-|\calB|)$.

Combing \eqref{sk}, \eqref{su} and \eqref{su:second}, it follows that
\begin{align}
\nn&\beta_\calB(\Phi_\mathrm{seq}^{\rmu}|P_\rmN,P_\rmA)\\*
\nn&\le {A_2}^2\left(
\frac{\exp\Big\{-(n-1)\Big(\Upsilon(P_\rmN,\lambda_2)-\frac{2|\calX|\log n}{n-1}\Big)\Big\}}{1-\exp\Big\{-\Big(\Upsilon(P_\rmN,\lambda_2)-\frac{2|\calX|\log n}{n-1}\Big)\Big\}}+\frac{\exp\Big\{-(n-1)\Big(\Omega(P_\rmN,P_\rmA,\lambda_1)-\frac{2|\calX|\log n}{n-1}\Big)\Big\}}{1-\exp\Big\{-\Big(\Omega(P_\rmN,P_\rmA,\lambda_1)-\frac{2|\calX|\log n}{n-1}\Big)\Big\}}\right.\\
&\left.\qquad+\frac{\exp\Big\{-(n-1)\Big(\Omega(P_\rmA,P_\rmN,\lambda_1)-\frac{2|\calX|\log n}{n-1}\Big)\Big\}}{1-\exp\Big\{-\Big(\Omega(P_\rmA,P_\rmN,\lambda_1)-\frac{2|\calX|\log n}{n-1}\Big)\Big\}}
+\frac{\exp\Big\{-(n-1)\Big(\Upsilon(P_\rmA,\lambda_2)-\frac{2|\calX|\log n}{n-1}\Big)\Big\}}{1-\exp\Big\{-\Big(\Upsilon(P_\rmA,\lambda_2)-\frac{2|\calX|\log n}{n-1}\Big)\Big\}}\right).
\end{align}
where $\Omega(P_\rmN,P_\rmA,\lambda)$ and $\Upsilon(P,\lambda)$ are defined in \eqref{omega} and \eqref{upsilon}, respectively.

Therefore, the misclassification exponent satisfies
\begin{align}
-\frac{1}{n}\log\beta_\calB(\Phi_\mathrm{seq}^{\rmu}|P_\rmN,P_\rmA)\ge
\min\big\{\Upsilon(P_\rmN,\lambda_2),~\Omega(P_\rmN,P_\rmA,\lambda_1),~
\Omega(P_\rmA,P_\rmN,\lambda_1),~\Upsilon(P_\rmA,\lambda_2)\big\}.\label{beta:unknown}
\end{align}

The proof of Theorem \ref{th:unknown} is now completed.

\subsection{Justification of the Benefit of Sequentiality with Unknown Number of Outliers}
\label{proof:sequentiality_unknown}

Recall the misclassification, false reject and false alarm exponents in Theorem \ref{fixed:unknown} for fixed-length test in Algorithm \ref{FA:unknown} and that in Theorem \ref{th:unknown} for sequential test in Algorithm \ref{CA:unknown}. Similarly to the case when the number of outliers is known, as discussed in \eqref{maxupsilon}, the Bayesian exponent is maximized when $\lambda_1\to 0$ and $\lambda_2\to f(P_\rmA,P_\rmN)$. In this case, we shall show the benefit of sequentiality when both tests achieve the same false alarm exponent.

Firstly, since the threshold $\lambda$ in the fixed-length test can be arbitrary, set the threshold $\lambda$ in Theorem \ref{fixed:unknown} as $\lambda_2$ for the sequential test in Theorem \ref{th:unknown}. It follows that the false alarm exponents in Theorems \ref{fixed:unknown} and \ref{th:unknown} are the same, i.e., $E_{\rm fa}(\Phi_\mathrm{seq}^{\rmu}|P_\rmN,P_\rmA)=E_{\rm fa}(\Phi_\mathrm{fix}^{\rmu}|P_\rmN,P_\rmA)$. Thus, we have shown that both tests achieve the same asymptotic performance under the null hypothesis.

We next show that the sequential test in Theorem \ref{th:unknown} achieves better performance under each non-null hypothesis than the fixed-length test in Theorem \ref{fixed:unknown}. Fix any $\calB\in\calS$. Since the exponent function $\Omega(P_1,P_2,\lambda)$ is non-increasing in $\lambda$ and $\lambda_1\le\lambda_2=\lambda$, it follows that $\max\{\Omega(P_\rmA,P_\rmN,\lambda_1),\Omega(P_\rmN,P_\rmA,\lambda_1)\}\ge\max\{\Omega(P_\rmA,P_\rmN,\lambda),\Omega(P_\rmN,P_\rmA,\lambda)\}$. Thus, the false reject exponents in Theorems \ref{fixed:unknown} and \ref{th:unknown} satisfy $E_{\zeta_\calB}(\Phi_\mathrm{seq}^{\rmu}|P_\rmN,P_\rmA)$ $\ge E_{\zeta_\calB}(\Phi_\mathrm{fix}^{\rmu}|P_\rmN,P_\rmA)$. As shown in Fig. \ref{fr_compare}, this inequality can be strict.

Finally, we show that the misclassification exponent for sequential test in Theorem \ref{th:unknown} when $\lambda_1\to 0$ and $\lambda_2\to f(P_\rmA,P_\rmN)$ is greater than that for fixed-length test in Theorem \ref{fixed:unknown}, i.e.,
\begin{align}
\nn&\min\big\{\eta(P_\rmA,P_\rmN),~\eta(P_\rmN,P_\rmA),~\gamma(P_\rmA,P_\rmN),~\gamma(P_\rmN,P_\rmA)\big\}\\*
&\le\min\big\{\Omega(P_\rmA,P_\rmN,0),~\Upsilon(P_\rmN,f(P_\rmA,P_\rmN)),~\Omega(P_\rmN,P_\rmA,0),~\Upsilon(P_\rmA,f(P_\rmA,P_\rmN))\big\}.\label{benefit:unknwon}
\end{align}
In \eqref{benefit:known}, we proved $\min\big\{\eta(P_\rmA,P_\rmN),~\eta(P_\rmN,P_\rmA)\big\}\le\min\big\{\Omega(P_\rmA,P_\rmN,0),~\Upsilon(P_\rmN,$ $f(P_\rmA,P_\rmN))\big\}$. Thus, it suffices to prove
\begin{align}\label{benefit:unknown1}
\min\big\{\gamma(P_\rmA,P_\rmN),~\gamma(P_\rmN,P_\rmA)\big\}\le\min\big\{\Omega(P_\rmN,P_\rmA,0),~\Upsilon(P_\rmA,f(P_\rmA,P_\rmN))\big\}.
\end{align}

\begin{itemize}
\item It follows from the definition of $\gamma(P_1,P_2)$ in \eqref{gamma} that
\begin{align}
\gamma(P_\rmN,P_\rmA)=\min_{\substack{(Q_1,Q_2,Q_3)\in\calP(\calX)^3:f(Q_1,Q_3)\le f(Q_1,Q_2)}}D(Q_1||P_\rmN)+D(Q_2||P_\rmN)+D(Q_3||P_\rmA).\label{gammaNA}
\end{align} It follows from the definition of $\Omega(P_1,P_2,\lambda)$ in \eqref{omega} that
\begin{align}\label{omegaNA}
\Omega(P_\rmN,P_\rmA,\lambda_1)&=\min_{\substack{(Q_1,Q_2,Q_3)\in\calP(\calX)^3:~f(Q_1,Q_3)\le\lambda_1}}D(Q_1\|P_\rmN)+D(Q_2\|P_\rmN)+D(Q_3\|P_\rmA).
\end{align}
Comparing \eqref{gammaNA} and \eqref{omegaNA}, we conclude that $\Omega(P_\rmN,P_\rmA,0)\ge \gamma(P_\rmN,P_\rmA)$ since $f(Q_1,Q_2)\ge 0$ for any pair of distributions $(Q_1,Q_2)\in\calP(\calX)^2$ while $\lambda_1\to 0$.
\item We next prove that $\Upsilon(P_\rmA,f(P_\rmA,P_\rmN))\ge\gamma(P_\rmA,P_\rmN)$. It follows from the definition of $\gamma(P_1,P_2)$ in \eqref{gamma} that
\begin{align}
\gamma(P_\rmA,P_\rmN)=\min_{\substack{(Q_1,Q_2,Q_3)\in\calP(\calX)^3:f(Q_1,Q_3)\le f(Q_1,Q_2)}}D(Q_1||P_\rmA)+D(Q_2||P_\rmA)+D(Q_3||P_\rmN).\label{gammaAN}
\end{align}
Setting $Q_3=P_\rmN$ in \eqref{gammaAN} and we obtain
\begin{align}\label{comp:gamma}
\gamma(P_\rmA,P_\rmN)\le\min_{\substack{(Q_1,Q_2,Q_3)\in\calP(\calX)^3:f(Q_1,Q_2)\ge f(Q_1,P_\rmN)}}D(Q_1||P_\rmA)+D(Q_2||P_\rmA).
\end{align}
Define the feasible region of $\Upsilon(P_\rmA,f(P_\rmA,P_\rmN))$ in \eqref{upsilon} and the right hand of \eqref{comp:gamma} as $\calF_\Upsilon:=\{(Q_1,Q_2)\in\calP(\calX)^2:~f(Q_1,Q_2)\ge f(P_\rmA,P_\rmN)\}$ and $\calG_\gamma:=\{(Q_1,Q_2)\in\calP(\calX)^2:~f(Q_1,Q_2)\ge f(Q_1,P_\rmN)\}$, respectively. Furthermore, define
\begin{align}\label{lambdaprime}
\lambda'(P_\rmN)=\min_{(Q_1,Q_2)\in\calG_\gamma}f(Q_1,P_\rmN).
\end{align}
To show $\Upsilon(P_\rmA,f(P_\rmA,P_\rmN))\ge\gamma(P_\rmA,P_\rmN)$, it suffice to prove that $\calF_\Upsilon\subseteq\calG_\gamma$, which is equivalent to $\lambda'(P_\rmN)\le f(P_\rmA,P_\rmN)$. Choosing $(Q_1,Q_2)=(P_\rmA,P_\rmN)$, the constraint of \eqref{lambdaprime} is satisfied and the objective function of \eqref{lambdaprime} equals $f(P_\rmA,P_\rmN)$. Thus, we have shown that $\lambda'(P_\rmN)\le f(P_\rmA,P_\rmN)$.
\end{itemize}
Therefore, the misclassification exponent for sequential test in Theorem \ref{th:unknown} when $\lambda_1\to 0$ and $\lambda_2\to f(P_\rmA,P_\rmN)$ is greater than that for fixed-length test in Theorem \ref{fixed:unknown}. As shown in Fig. \ref{mis_compare}, the benefit can be strict.

Since all exponents of misclassification, false reject and false alarm probabilities for the sequential test in Theorem \ref{th:unknown} are greater than or equal to that for the fixed-length test in Theorem \ref{fixed:unknown}, the benefit of sequentiality in terms of Bayesian error exponent naturally holds.

\bibliographystyle{IEEEtran}
\bibliography{IEEEfull_paper}

\end{document}